\documentclass{egpubl}
\usepackage{eurovis2026}
\usepackage{lipsum}
\usepackage{caption}
\usepackage{microtype}
\usepackage{subcaption}
\usepackage{booktabs}
\usepackage{amsfonts}
\usepackage{amssymb}
\usepackage{amsmath}
\usepackage{float}
\usepackage{color}
\EuroVisShort  

\usepackage[T1]{fontenc}
\usepackage{dfadobe}

\usepackage{cite}  
\BibtexOrBiblatex
\electronicVersion
\PrintedOrElectronic
\ifpdf \usepackage[pdftex]{graphicx} \pdfcompresslevel=9
\else \usepackage[dvips]{graphicx} \fi

\usepackage{egweblnk}

\title[ShapDBM]{ShapDBM: Exploring Decision Boundary Maps in Shapley Space}

\author[Watkin et al.]{
    \parbox{\textwidth}{
    \centering
    Luke Watkin$^{1}$\orcid{0009-0003-5407-9359}, Daniel Archambault$^{1}$\orcid{0000-0003-4978-8479}, Alex Telea$^{2}$\orcid{0000-0003-0750-0502} 
    }
    \\
    {\parbox{\textwidth}{
        \centering 
        $^1$School of Computing, Newcastle University, UK\\
        $^2$Department of Information and Computing Science, Utrecht University, Netherlands
        }
    }
}

\begin{document}
    \teaser{
        \vspace{-0.15cm}
        \centering
        \includegraphics[width=0.9\linewidth]{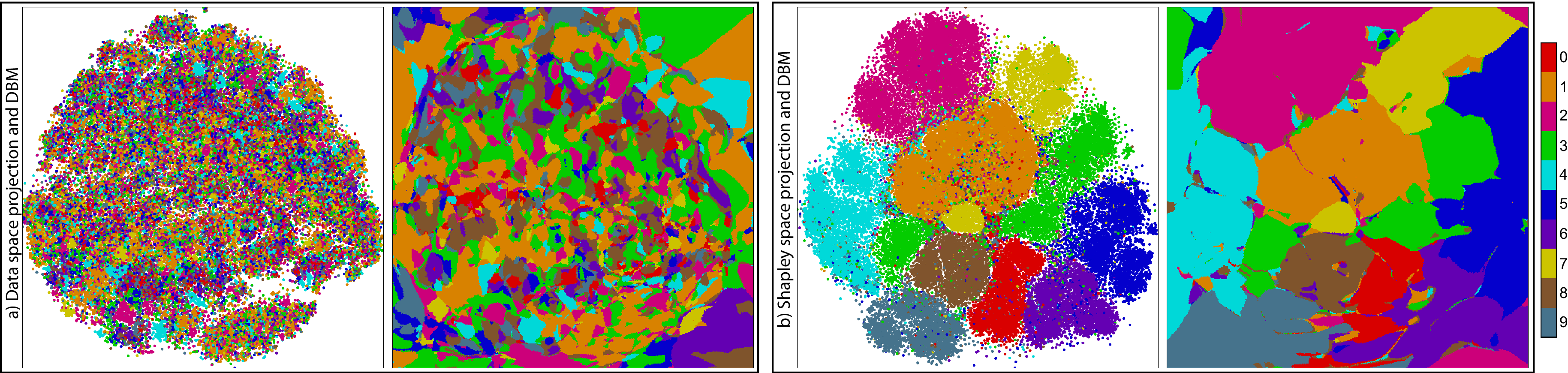}
        \caption{Decision Boundary Maps (DBM) for SVHN created using dimensionality reduction from (a) data space $P(D)$ and (b) Shapley space $P(S)$. The data space DBM is fragmented into small zones, falsely suggesting an issue with the classifier, and also being useless for any practical task. Our method creates a less fragmented, more interpretable result that also agrees far better with model performance.}
        \label{fig:svhn_data_shap_maps}
    }
    
    \maketitle
    
    \begin{abstract}
        Decision Boundary Maps (DBMs) are an effective tool for visualising machine learning classification boundaries. Yet, DBM quality strongly depends on the dimensionality reduction (DR) technique and high dimensional space used for the data points. For complex ML data, DR can create many mixed classes which yield DBMs that are hard to use or even misleading. We propose a new technique to compute DBMs by transforming data space into Shapley space and computing DR on it. Compared to DBMs computed directly from data, our maps have similar or higher quality metric values and visibly more compact, easier to explore, decision zones that better agree with measured model performance.

    \end{abstract}  

    \section{Introduction}
        Decision Boundary Maps (DBMs)\,\cite{dbm_original} show how a trained classification model partitions its high-dimensional data space. 
        They use dimensionality reduction (DR) to project data to 2D, train an inverse projection to reverse this mapping to create samples for all image pixels, and colour pixels to map the model's decision at these samples. Recent work has shown that all existing \emph{data space} DBMs -- which map samples from data space to 2D via DR -- only depict a fixed/smooth interpolating \emph{surface} through these samples\,\cite{wang_2024}. Complex ML models have complex decision zones. Given the inherent limitation of all DR methods in preserving neighborhoods\,\cite{espadoto19}, this leads to poor quality, fragmented, DBMs which wrongly convey that the trained model (and not the DBM) is problematic\,\cite{rodrigues19, DBM_eval}.
        
        We show that \emph{transforming} the data before projection can lead to higher-quality, easier to interpret, DBMs, by \textbf{ShapDBM}, a technique that creates DBMs using the data's Shapley values\,\cite{shap}, a feature importance metric known to produce projection scatterplots with greater clarity\,\cite{shap_clustering}. Figure~\ref{fig:svhn_data_shap_maps} shows a high quality map (created by ShapDBM) of the full SVHN dataset for the first time in the literature. On three case studies, we show that ShapDBM delivers DBMs which are better visually and \emph{vs} established quality metrics than existing DBMs using data-space projections. 

\vspace{-0.15cm}
\section{Background and Related Work}
    \label{sec:related_work}
    \textbf{Preliminaries:} 
        Let $D \subset \mathbb{R}^n$ be a \emph{training} dataset with samples $(x_i^1, \ldots, x_i^N) \in \mathbb{R}^n, 1 \leq i \leq N$ with labels $y_i \in \{1, \dots, C\}$. 
         $D_{test} \subset \mathbb{R}^n$ is a \emph{test} dataset, $D \cap D_{test} = \varnothing$. A classifier $f$ trained on $D$ does the mapping $f:\mathbb{R}^n \rightarrow C$ so that $f(x_i) = y_i$ for all $x_i \in D \cup D_{test}$. A dimensionality reduction or \emph{projection} $P$ performs a mapping $P : D \rightarrow \mathbb{R}^q$ ($q = 2$ for our purposes)\,\cite{nonato18,espadoto19}. 
        We denote the use of $P$ on a dataset $D$ as $P(D)$.
        An \emph{inverse projection} $P^{-1} : \mathbb{R}^q \rightarrow \mathbb{R}^n$ aims to approximately revert the effect of $P$\,\cite{iLamp,NNINV,DBM_eval} so that $P^{-1}(P(x)) \approx x, x \in D$. Once trained, $P^{-1}$ can be applied to \emph{any} point in $\mathbb{R}^q$. When used on points outside $P(D)$, $P^{-1}$ creates new synthetic samples consistent with $D$'s distribution.
                 
    \noindent\textbf{Feature Importance:}
        One way to study the behaviour of ML models is to examine which features (dimensions of a sample $x \in D$) have the highest contribution to the output. \emph{Model specific} approaches~\cite{GRAD_CAM,deconv_networks} 
        explain a specific architecture. \emph{Model agnostic} approaches\,\cite{ANCHORS, LIME, shap, lrp} handle any ML model. 
        LIME \cite{LIME} builds a \emph{local} linear model to explain the neighbourhood of a specific input, so cannot offer global explanations.
        ShapDBM uses Shapley values\,\cite{shap}, a global  feature importance metric. Samples treated the same by the model should have similar Shapley values. 
        Originating in game theory\,\cite{shap_original}, the Shapley value of a feature $1 \leq k \leq n$ of a sample $x$ is computed by examining the 
        difference in outcome $f(x')$, \emph{e.g.}, prediction confidence for classifiers, for subsets $x' \in (x^1, \ldots x^N)$ where $x^k$ is present \emph{vs} subsets where $x^k$ isn't. 
        As exact computation is exponential in $N$,  approximation methods have been proposed, such as Monte-Carlo sampling\,\cite{MONTE_CARLO_SHAP} and architecture-specific approaches \emph{e.g.} DeepExplainer\,\cite{DEEPEXPLAINER} -- the latter being what we use in our work.

    \noindent\textbf{Decision Boundary Maps (DBMs):}
        A DBM\,\cite{dbm_original, supervised_dbm, fastmap, Deepview} is a 2D image depicting the output of a classifier $f$. Given a projection $P(D)$ of a dataset $D$, one first trains an inverse projection $P^{-1}$. Next,  pixels $y$ of a given regular 2D grid (image) are coloured by $f(P^{-1}(y))$. Compared to traditional scatterplots $P(D)$\,\cite{PROJECTIONS_PAPER},
        DBMs show $f$'s behaviour over a \emph{compact} space -- the image -- and reveal \emph{decision zones} (where $f$ has the same outcome) and their separating \emph{decision boundaries}.

\vspace{-0.15cm}
\section{Method}
    \label{sec:method}
    As existing DBM methods, we  start with a dataset $D$ and a trained classifier $f$ and work as follows:

    \noindent\textbf{1. Compute Shapley values and their projections:} 
        To obtain well-separated decision zones in the DBM, we aim to create projections with well-separated point clusters. Depending on the ML model, samples $x$ far apart in $\mathbb{R}^n$, \emph{e.g.}, two different colour images of digit `2' may yield the same class $f(x)$. Yet, samples with similar influential features have similar Shapley value-sets and also land in the \emph{same decision zone}.
        We compute a Shapley value set $S \subset \mathbb{R}^n$ for all samples $D$ and project $S$ using any user-chosen $P$ to get $P(S)$.

    \noindent\textbf{2. Inverse projection:} 
        We train our inverse projection $P^{-1}$ by minimising the error $\| P^{-1}(P(x)) - x \|$ over $D$. As $P^{-1}$ maps to $D$ (not $S$) so  we can use $P^{-1}$ to create synthetic samples for $f$.
            
    \noindent\textbf{3. Create 2D point grid:} 
        We create a 2D uniform grid $I$ of $r^2$ pixels (user-selected resolution $r$) covering the bounds of $P(S)$. In each pixel $p$, we pick $k=\{1..l\}$ random locations to reduce sampling artifacts\,\cite{dbm_original}, yielding a set $M(p,k), p\in I, k \in \{1..l\}$.
    
    \noindent\textbf{4. Colour pixels by predictions:} 
        We use $P^{-1}$ to map each $M$ point to a synthetic sample in $\mathbb{R}^n$ compatible with the model $f$. We colour $p$ by the most common class of its samples $M(p,k)$; and optionally map this class frequency\,\cite{dbm_original} or the distance of $P^{-1}(p)$ to its closest decision boundary in $\mathbb{R}^n$\,\cite{Machado2024} 
        to saturation in $I$. 
        
\section{Experiment Setup}
\noindent\textbf{Projections and Inverse Projections:} We use t-SNE\,\cite{tsne}; and UMAP\,\cite{umap} for $P$ and NNInv\,\cite{NNINV} (a neural network based approach) for $P^{-1}$ as an earlier evaluation\,\cite{DBM_eval} found that UMAP and t-SNE are optimal pairings for NNInv. We next focus on t-SNE and show UMAP results in the supp. material.
     
\noindent\textbf{Datasets:} We use three classification datasets of increasing difficulty. We start with MNIST\,\cite{MNIST} (70k images of digits 0-9), a well-known dataset used in previous DBM works\,\cite{supervised_dbm, dbm_original, fastmap}, making it ideal as  baseline.
Next, we use SVHN\,\cite{SVHN} (almost 100k full-colour images). To our knowledge, this is the first time that high-quality DBMs of SVHN have been produced. Finally, we use the CIFAR-4 subset of CIFAR-10\,\cite{CIFAR10} that includes all samples from classes \emph{airplane}, \emph{cat}, \emph{deer} and \emph{ship} (24k full-colour images). 
Initial experiments on CIFAR-10 produced DBMs with $< 20\%$ accuracy. We use a CIFAR-4 to reduce the difficulty, leaving CIFAR-10 for future work.
Note that except for DeepView\,\cite{Deepview}, which is a very slow method, no DBM has tackled the full CIFAR-10. For stability and efficient computation of Shapley values, we downsample SVHN and CIFAR-4 to  $28^2$ pixels.
       
\noindent\textbf{Model:} 
We use a Convolutional Neural Network (CNN) for $f$. While working well on our case studies (Tab.~\ref{tab:map_accuracy_precision_recall}), this is irrelevant for our added-value claim: DBMs are most useful to explore
underperforming models to improve them (see Sec.~\ref{sec:insights}). Our full code and datasets are openly available\,\cite{shapdbm}.

    \noindent\textbf{Hyperparameters:} 
        We fix a random seed for reproducibility; use the default t-SNE parameters from scikit-learn; and set map resolution $r = 500$ pixels and $l = 1$ random samples/pixel. We estimate Shapley values with DeepExplainer with 100 test samples to integrate over, following its documentation.

\vspace{-0.15cm}        
\section{Results}
    \subsection{Boundary Map Quality}
        We use adapted ML quality metrics to gauge map accuracy ($MA$), map precision ($MP$), and map recall ($MR$), all computed with predicted labels. All metrics need samples from $D$ so they consider only pixels covered by $P(D)$ or $P(S)$.
        $MA$ measures how well a DBM fits the data (how many $D$ samples are in the correct decision zone)\,\cite{DBM_eval}). 
        To examine a given class $c$, we adapt precision and recall: Map precision ($MP$) measures how many samples labelled $c$ lie on a pixel predicted as $c$ (low $MP$ means the map \emph{over-represents} $c$). Map recall ($MR$) measures how well a map finds all pixels predicted as $c$ (low $MR$ tells that $c$ is \emph{under-represented}). For brevity, we show averages of $MP$ and $MR$ over all $c$ in Tab.~\ref{tab:map_accuracy_precision_recall}.

\begin{table}[h]
\caption{Model accuracy (Acc), average precision (AP), average recall (AR). Maps  average accuracy (MA), average precision (MP), and average recall (MR) scores,  variance shown in brackets. }
\label{tab:map_accuracy_precision_recall}
\centering
\scriptsize
\setlength{\tabcolsep}{2.5pt}
\begin{tabular}{l ccc ccc ccc}
\toprule
& \multicolumn{3}{c}{Model} & \multicolumn{3}{c}{Data} & \multicolumn{3}{c}{Shapley} \\
Dataset & Acc & AP & AR & MA & MP & MR & MA & MP & MR \\
\midrule
MNIST   & 99.6 & .996 & .995 & 96.8 & .97(.0001) & .97(.0003) & 91.8 & .93(.01)  & .92(.004) \\
SVHN    & 95.3 & .950 & .949 & 25.3 & .27(.02)   & .22(.02)   & 85.4 & .86(.005) & .83(.005) \\
CIFAR-4 & 89.9 & .899 & .899 & 49.4 & .51(.002)  & .49(.06)   & 44.4 & .63(.08)  & .44(.13) \\
\bottomrule
\end{tabular}
\vspace{-0.2cm}
\end{table}
    
        \begin{figure*}[htb]
            \centering
            \includegraphics[width=0.8\linewidth]{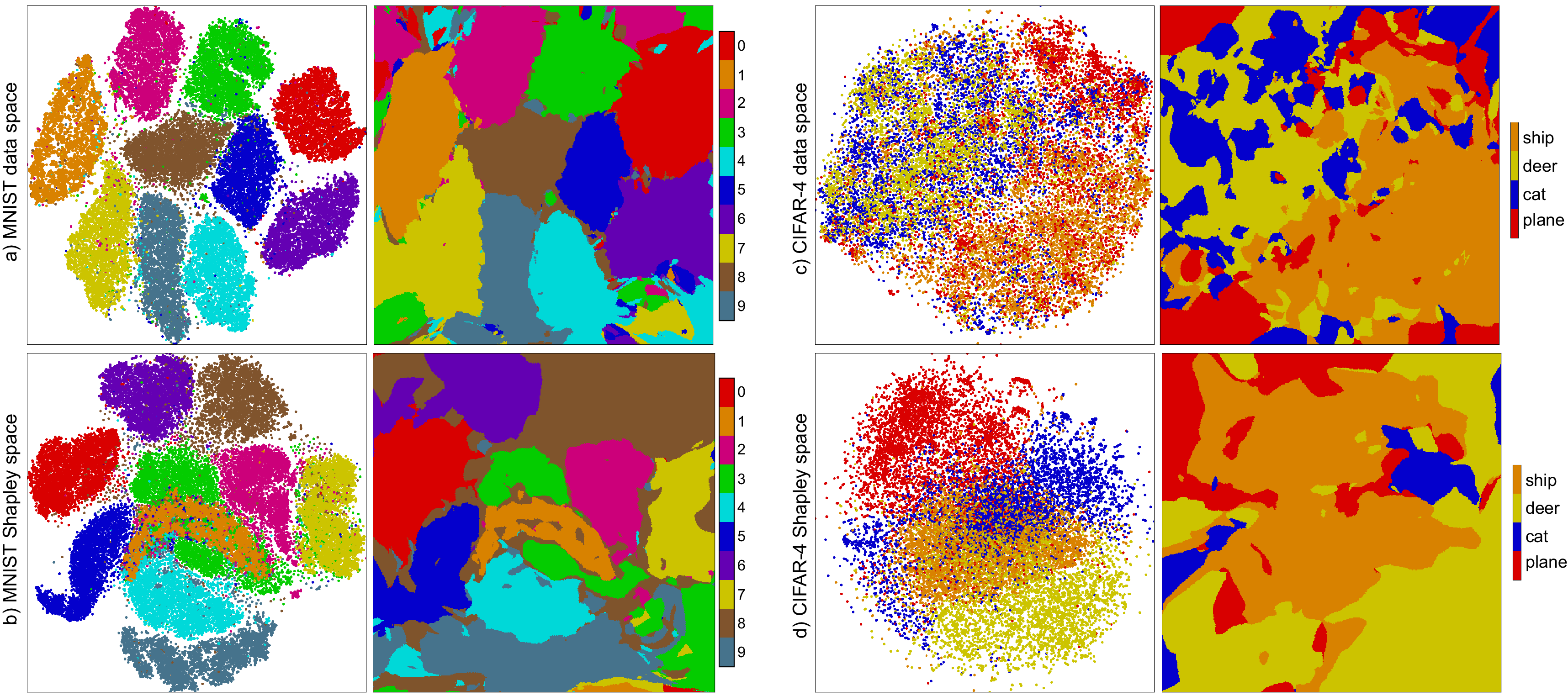}
            \caption{Data space projections $P(D)$ and DBMs (a,c) and Shapley space projection $P(S)$ and DBMs (b,d), MNIST and CIFAR-4.}
            \label{fig:mnist_cifar_4}
            \vspace{-0.2cm}
        \end{figure*}
        
        \noindent\textbf{MNIST:}
            Figure~\ref{fig:mnist_cifar_4}a,b shows the data space and Shapley value DBMs for MNIST. 
            Both maps show well-separated decision zones for each class. Data-space DBM (Fig.~\ref{fig:mnist_cifar_4}a) has a marginally higher $MA = 95.8\%$ compared to our Shapley-DBM ($MA = 91.8\%$). 
            We attribute this to a higher number of `disputed' regions visible on the data-space scatterplot between class 1 (orange) and 4 (light blue) and class 3 points (green) in class 1.   
            Such regions become class 8 (brown) in the DBM, reflected in a relatively low $MR_3=0.76$ and $MP_8=0.66$ for the Shapley DBM, telling that this map struggles to locate all class 3 regions.
            The Shapley DBM is made of mainly class 8 (brown) pixels which appear in regions between known clusters of points. We also see this in the data-space DBM but to a lesser degree. 
            Overall, while the data-space DBM scores higher than our method for MNIST, our method still yields a DBM of comparable quality. 
    
        \noindent\textbf{SVHN:}
            Figure~\ref{fig:svhn_data_shap_maps}b shows our Shapley-based DBM. 
            Compared to the data-space DBM (Fig.~\ref{fig:svhn_data_shap_maps}a), we see much better separated, less fragmented, decision zones, reflected in $MA$ increasing from $25.3\%$ to $85.4\%$. The simple CNN model used to classify SVHN should generate compact decision zones with smooth boundaries\,\cite{montufar14,alfarra22} so our Shapley DBMs better capture this than data-space ones.
            Inter-class boundaries in the data-space DBM show only small islands matching point clusters in the scatterplot. Class 1 (orange) occupies a large fraction of the map while $MR_1 = 0.52$ and $MP_1 = 0.4$. 
            In contrast, $MP_1=0.78$ and $MR_1=0.97$ for our Shapley-based DBM, telling that it has significantly higher quality. 
            Yet, ShapDBM is not artifact free: small islands appear near the class 1 (orange) decision zone, where we see disputed regions on boundaries with classes 7 (yellow) and 3 (green). Several also exist in the far corners of the map, where DBMs are known to be inaccurate\,\cite{DBM_eval}.
            Overall, for SVHN, we find that our method produces significantly higher quality DBMs that data-space DBMs.
        
        \noindent\textbf{CIFAR-4:}
            The CIFAR-4 difficulty is shown by the quality of both data-space and our DBMs (Fig.~\ref{fig:mnist_cifar_4}c,d). Both struggle to get $> 50\%$ $MA$. 
            Data space performs slightly better ($MA=49.4\%$) than Shapley space ($MA=44.4\%$).
            Both maps struggle detecting \emph{airplanes} (red), with low $MR$ scores of $0.19$ (data space) and $0.16$ (our method). Decision zones for \emph{airplanes} are often misclassified as \emph{ships} (orange) -- see the large \emph{airplane} points cluster in Fig.~\ref{fig:mnist_cifar_4}d which do not map to a decision zone for the class, adding to the low $MP$ score. \emph{ship} has the highest $MR$ in both maps 
            (data space: $0.78$; our method: $0.94$); yet, both maps have low $MP$ scores for \emph{ship} (data space: $0.45$; our method: $0.33$), telling over-representation. In projections, we see clearer class separation with the projected Shapley values (Fig.~\ref{fig:mnist_cifar_4}d), especially between classes \emph{deer} (yellow) and \emph{cat} (blue), which form a visible confusion zone in the data-space projection (Fig.~\ref{fig:mnist_cifar_4}c). In contrast, our method creates far better separated same-class samples in the projection, leading to a far less fragmented DBM.
            Overall, while neither of the two DBMs is close to ideal, our method creates a better map in terms of class and decision zone delineation.
        
        \vspace{-0.15cm}
        \subsection{Inverse Projection Accuracy}
            We study how well $P^{-1}$ (NNInv in our case) reconstructs $D$ samples from $P(D)$ (data space) and $P(S)$ (Shapley Space) based on a round trip $P \rightarrow P^{-1}$
            as in prior work\,\cite{dbm_original,unprojection,DBM_eval} (see supp. material for details).
                            
                            
            \noindent\textbf{SVHN:}
                $P(D)$ can recover colour well but less so structure.
                Shapley space derived samples consistently have muted colours but better structure. 
                $P(D)$ reconstructions make structure errors, such as the 2 incorrectly recovered as a 3. 
                We hypothesize that $P(D)$ creates small clusters due to colour; $P(S)$ considers colour less, being driven by how $f$ treats samples, so $P^{-1}$ recovers colour less than structure. 
            
            \noindent\textbf{CIFAR-4:}
                Both $P(D)$ and $P(S)$  reconstructions do not resemble the original images. $P(D)$ yields slightly more faithful results that keep average colour though the actual shape is missing. 
                We see similar behaviour to SVHN in the $P(S)$ reconstructions, supporting our earlier stated hypothesis.
                The most accurate reconstructions belong to class \emph{deer} (first, sixth columns), where we see a faint shape. 

                The fact that $P(S)$ cannot reconstruct samples $D$ as well as $P(D)$ is not an issue: The key aim of DBMs is
                not to capture the \emph{data} $D$ but the \emph{behaviour} of a trained model $f$ on $D$. Simply put, the poor reconstructions we see mean that our DBMs sample the data space farther from $D$, which is good for exploring $f$; in contrast, $P(D)$ samples tightly close to $D$, which is less informative\,\cite{wang_2024}. 
                Our mapping $D \rightarrow S$ does not cause information loss -- it actually creates \emph{more valid} samples that $f$ should be able to classify.

        \subsection{Actionable Insights Provided by Shapley DBMs}
        \label{sec:insights}

            \begin{figure}
                \centering
                \includegraphics[width=0.75\linewidth]{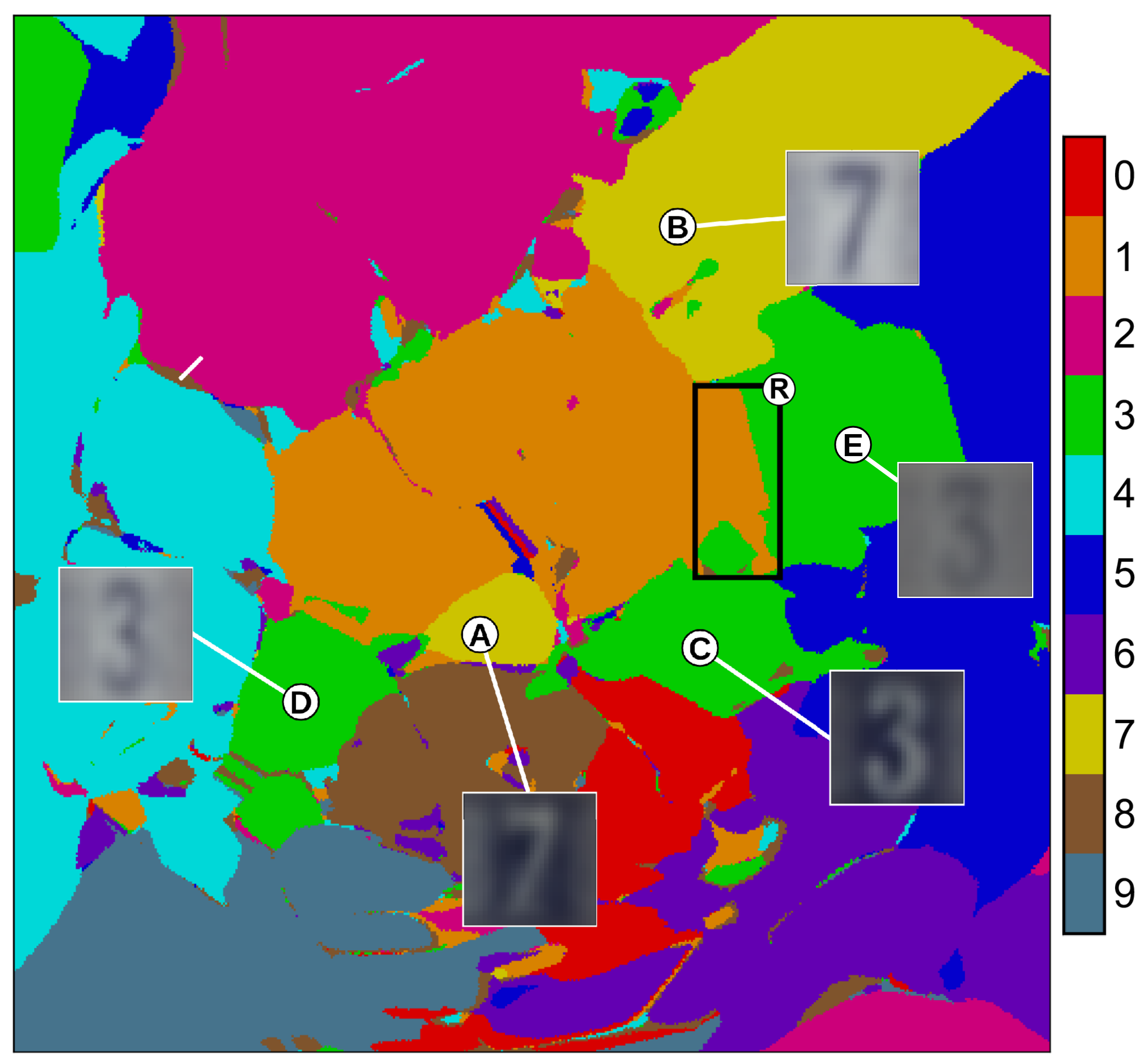}
                \caption{Shapley DBM for SVHN with samples A-E from classes having multiple decision zones. A region R is marked where boundaries strongly differ from what the projection plot (Fig.~\ref{fig:svhn_data_shap_maps}b) conveys.}
                \label{fig:svhn_illustrated}
                \vspace{-0.25cm}
            \end{figure}

            \noindent\textbf{What Shapley DBMs tell:} In the plot $P(S)$ (Fig.~\ref{fig:svhn_data_shap_maps}b), we see a large class 3 (green) band right to class 1 (orange) which suggests a solid, compact, class 1 decision zone here. Yet, the DBM shows a different story - the band is narrowed by the class 1 (orange) zone, see Fig.~\ref{fig:svhn_illustrated}(R). We see the same discrepancy between the plot $P(S)$ and DBM in many other regions -- overall, the DBM shows that decision zones are far less simple and compact than the plot suggests. This is a first important take-away message: raw projection plots can convey 
            \emph{misleading} insights on how a model operates, mainly due to point  overplotting and unspecified drawing order; Shapley DBMs fix this since every pixel has a single decision assigned to it.

            Let us refine this analysis: Figure~\ref{fig:svhn_data_shap_maps}b shows several disconnected decision zones for classes 7 (yellow) and 3 (green). We brush these zones and see that they correspond to bright digits on dark background and dark digits on bright background (Fig.~\ref{fig:svhn_illustrated}A-E). In fact, class 3 (green) has three such zones which corresponds to light, grey and dark. The scatterplot representation indicates only two such zones. This was previously noted by Rauber et al.\,\cite{PROJECTIONS_PAPER} on SVHN but only for a \emph{small} subset of this dataset -- not the full dataset as in this work. The data space DBM (Fig.~\ref{fig:svhn_data_shap_maps}a) also shows many zones for a class but its huge fragmentation makes it nearly impossible to use: we cannot brush every single island to understand what makes those samples differ from other same-class samples.
            
            \noindent\textbf{Improving Model Performance:} Using the above insight, we train a CNN on SVHN with a Sobel filter applied to classes 3 and 7 that removes brightness and only keeps the digits' edges in the input images (see supp. material for details). We get $1.3\%$ accuracy gain, double the $0.65\%$ accuracy gain reported by Rauber et al. who used the same CNN model but did Sobel filtering on \emph{all} classes. Our Shapley DBMs allowed us to find the precise classes get split into distinct decision zones to preprocess \emph{only} those images. Rauber et al. assessed the model using projection plots (and not DBMs) which, as we saw above, can be misleading due to overplotting and drawing order; and thus wrongly inferred this split to occur for all classes.

    \vspace{-0.15cm}
    \section{Discussion}
        \label{sec:discussion}
        \noindent\textbf{Scalability:}
            Computing Shapley values is costly (times, all case studies, in supp. material): $\approx21$ hours on our largest dataset (SVHN); $\approx5$ hours on our smallest dataset (CIFAR-4). Subsets can yield a high $MA$ -- a 10k SVHN subset yields a map with $MA=86.2\%$ in $\approx3$ hours (see supp. material). While our cost largely exceeds existing DBM methods (except DeepView\,\cite{Deepview} which times similarly), this is due to Shapley value computation being largely CPU-bound on large datasets for stability reasons.
            Our method yields much higher-quality DBMs (lower measured errors and less visual fragmentation), both key for correctly assessing how a ML model works, our core goal. Optimizing speed is left to future work.
        
        \noindent\textbf{Validity:}
            Using $P(D)$ on complex datasets creates fragmented DBMs that falsely imply issues with the model $f$. $P(S)$ represents how $f$ views the dataset $D$, grouping distant data samples (which $f$ treats similarly) in Shapley space. 
            DBMs are a fixed surface cutting $\mathbb{R}^n$ that shows how decision boundaries meet between given points\,\cite{DBM_eval}. In our case, this surface is mainly determined by the choice of $P$ and $P^{-1}$. The transformation $D \rightarrow S$ changes the position of point clusters in $\mathbb{R}^n$ and as such yields a different, but by construction valid, DBM.
            Interpolation in $P(D)$ is fixed between known samples for DBMs. In contrast, $P(S)$ places the manifold outside these regions in data space, thus covering $f$ better. 

        \noindent\textbf{Future Work:} 
            NNInv inverse projection quality could be improved by deconvolution operations\,\cite{deconv_networks}. Aggregation\,\cite{grouseflocks, antichaincikm} could improve visual quality via multilevel approaches. 
            Our approach focuses on Shapley values as a transformation to improve DBM quality, yet other approaches may yield similar quality improvements with less computational effort.

\vspace{-0.15cm}        
\section{Conclusion}
        We presented ShapDBM, a method that creates high-quality DBMs using projected Shapley values.  
        Our method achieves similar or higher quality on three datasets \emph{vs} standard data-space DBMs. Our work shows that data space \emph{transforms} (by Shapley and possibly other methods) yield better maps and that such transforms deliver actionable insights that ML engineers can take from DBMs.

\vspace{-0.15cm}        
\section*{Acknowledgements} For open access, we have applied a Creative Commons Attribution (CC-BY) license to any Author Accepted Manuscript arising from this paper. 
        
\bibliographystyle{eg-alpha-doi}
\bibliography{egbibsample}
\newpage

\title[Supplementary Material]{ShapDBM: Visually Scalable Decision Boundary Maps\\ in Shapley Space}

\section*{Supplementary Material}
\maketitle

\section{MNIST Inverse Projection Accuracy}
    \textbf{Context:} In section 5.2, we examine how NNInv reconstructs samples, based on projections in data space and Shapley space. The following section is our discussion for the MNIST case study, which was removed due to space constraints.
    
    \begin{figure}[H]
        \centering
        \includegraphics[width=1.0\linewidth]{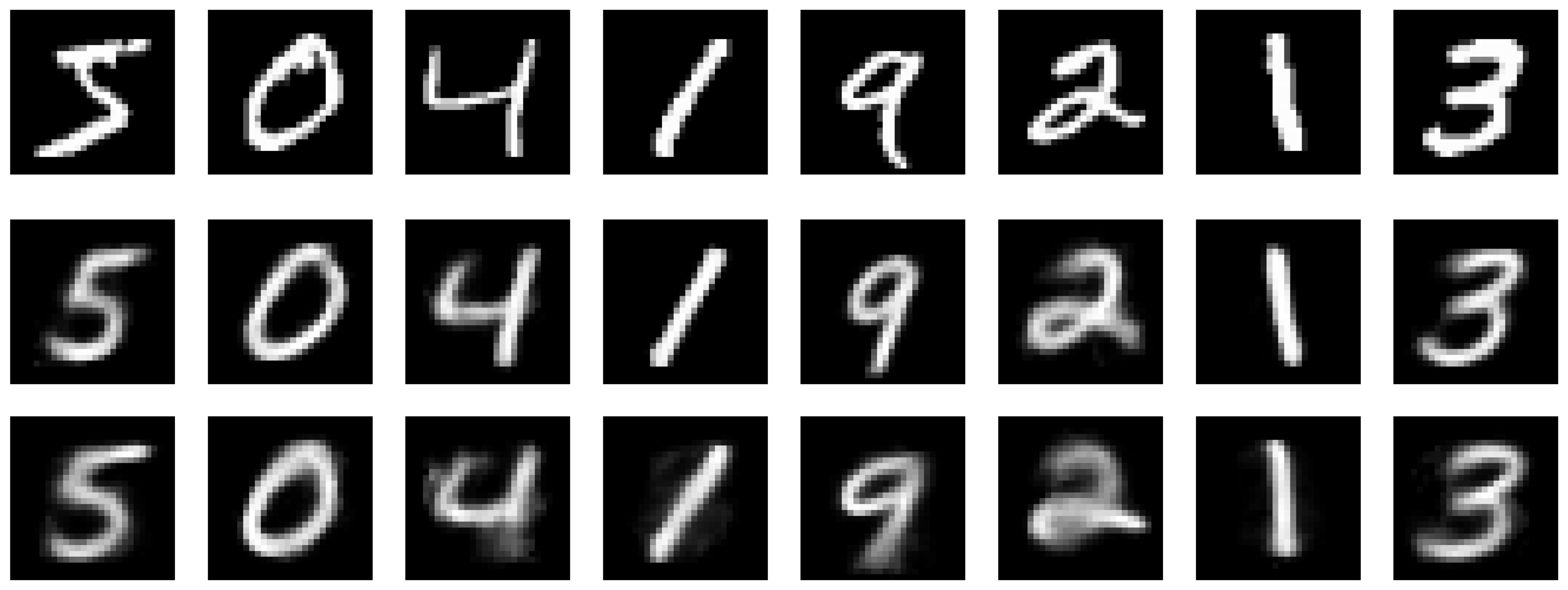}
        \caption{Samples from MNIST dataset (top) and reconstructed samples based on projections using NNInv trained on data space projections (middle) or projected Shapley values (bottom)}
        \label{fig:MNIST_inverse_grid}
    \end{figure}
    
    Figure \ref{fig:MNIST_inverse_grid} shows samples from the MNIST dataset (top row) and reconstructions based on their projected samples in data space (middle row) and Shapley value space (bottom row). In both cases, the projected samples are structurally the same but noticeably less sharp. Consider the class 4 sample (Figure \ref{fig:MNIST_inverse_grid} column three) where the "tail" 
   of the Shapley-derived sample is significantly blurrier than both the original sample and the data-derived sample. We observe similar behaviour with the class 2 sample (Figure \ref{fig:MNIST_inverse_grid} column six) where the hole of the loop has been filled. We attribute this behaviour to the Shapley value projections preserving parts of the samples that are the most influential to the classification.
   
\section{Inverse Projection Accuracy Samples}
\begin{figure}[H]
        \centering
        \includegraphics[width=0.9\linewidth]{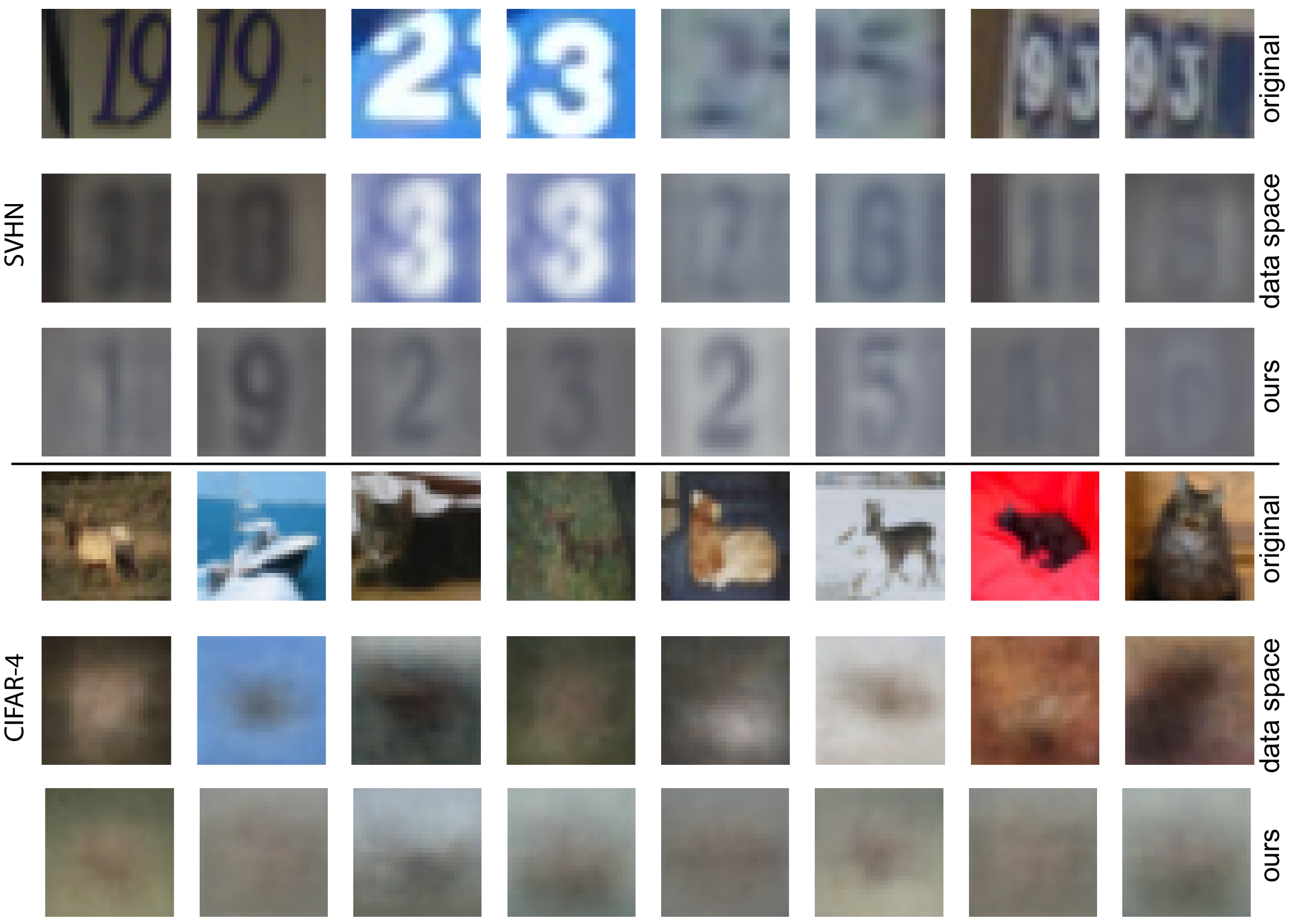}
        \caption{SVHN and CIFAR-4 samples reconstructed using NNInv from data space and Shapley space projections.}
        \label{fig:invproj}
        \vspace{-0.15cm}
    \end{figure}

    \textbf{Context:} In section 5.2, we examine how our chosen inverse ($P^{-1}$) method can reconstruct known samples from SVHN and CIFAR-4, based on one round trip of $P \rightarrow P^{-1}$. Fig. \ref{fig:invproj} presents the reconstructions based on data space projections and Shapley space projections for comparison against the original samples.

\section{SobelSVHN Samples}
\begin{figure}[H]
        \centering
        \begin{subfigure}[t]{\linewidth}
            \centering
            \includegraphics[width=0.75\linewidth]{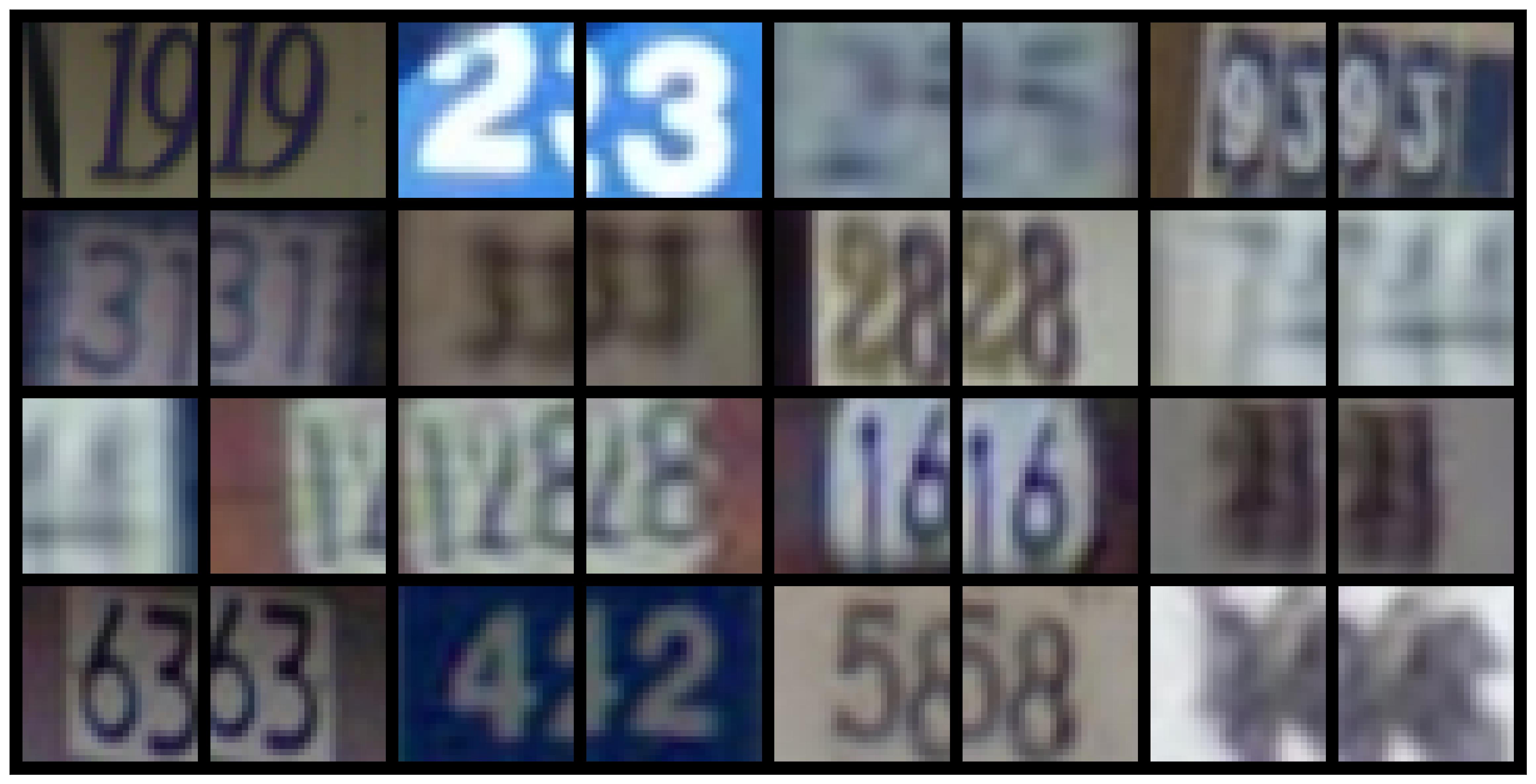}
            \caption{Resized SVHN.} 
            \label{subfig:svhn}
        \end{subfigure}
        \begin{subfigure}[t]{\linewidth}
            \centering
            \includegraphics[width=0.75\linewidth]{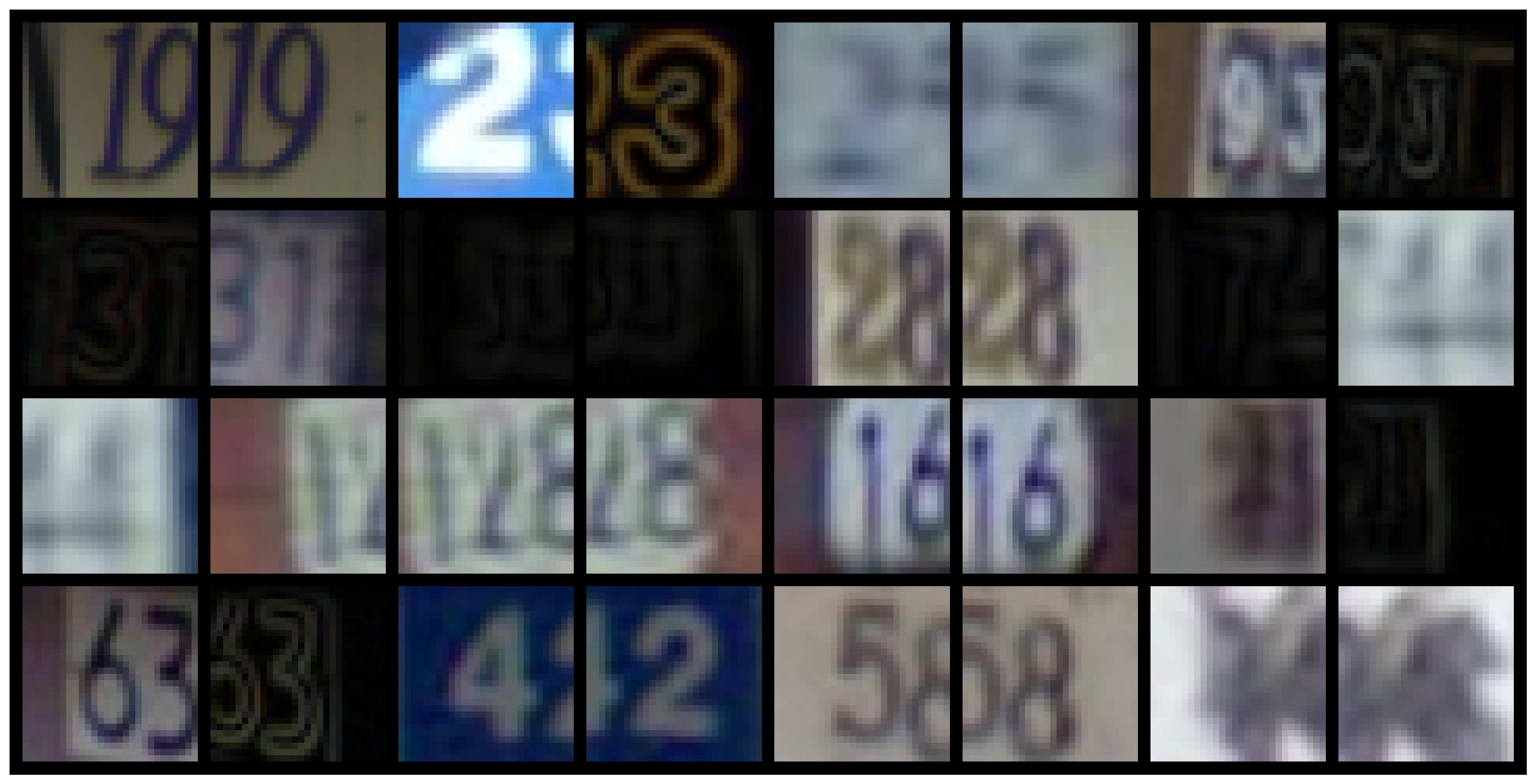}
            \caption{Resized SVHN with Sobel filter applied to samples from classes 3 and 7.} 
            \label{subfig:sobel_svhn}
        \end{subfigure}
        \caption{(a) Samples taken from the SVHN dataset used in our case study (note that the samples are resized to 28 pixels instead of 32) and (b) a variant of our SVHN case study created by passing samples of classes 3 and 7 through a Sobel filter.}
        \label{fig:sobel_svhn}
        \vspace{-0.5cm}
    \end{figure}

    \textbf{Context:} In section 5.3, we examine how ShapDBMs can provide actionable insights into classifier behaviour to improve performance. In our example, we apply a Sobel filter to samples from classes 3 (green) and 7 (yellow), as we noted distinct decision zones corresponding to variations in the input data in the final DBM, to create a variant of SVHN on which to train a classifer (hence referred to as \emph{SobelSVHN}). Figure \ref{fig:sobel_svhn} shows a sample batch taken from SVHN (Fig. \ref{subfig:svhn}) and the corresponding batch from the \emph{SobelSVHN} dataset (Fig. \ref{subfig:sobel_svhn}).

\section{Shapley Value Estimation Time}
    \textbf{Context:} In section 6, we discuss the computational cost of Shapley values for our case studies. Table \ref{tab:shap} shows the time required for our trained CNN to estimate Shapley values for all samples across all studies. See section 6 for related discussion. 
    \begin{table}[H]
        \caption{Computation times for Shapley values for all case study datasets, in relation to the size of the dataset (N), the size of their training sets on which Shapley values are calculated ($N_{train}$) and dimensionality (n).}
         \label{tab:shap}
        \centering
        \small
        \begin{tabular}{| l | r r r r |}
            \hline
            Dataset & $N$ & $N_{train}$ & $n$ & Estimation Time (hours)\\
            \hline
            MNIST   & 70k & 60k  & 784  & 16 \\
            SVHN    & 99k & 73k  & 2352 & 21 \\
            CIFAR-4 & 24k & 20k  & 2352 & 5  \\
            \hline
        \end{tabular}
    \end{table}\newpage

\section{SVHN Subset}
\begin{figure}[H]
        \centering
        \begin{subfigure}[t]{\linewidth}
            \centering
             \includegraphics[width=0.45\linewidth]{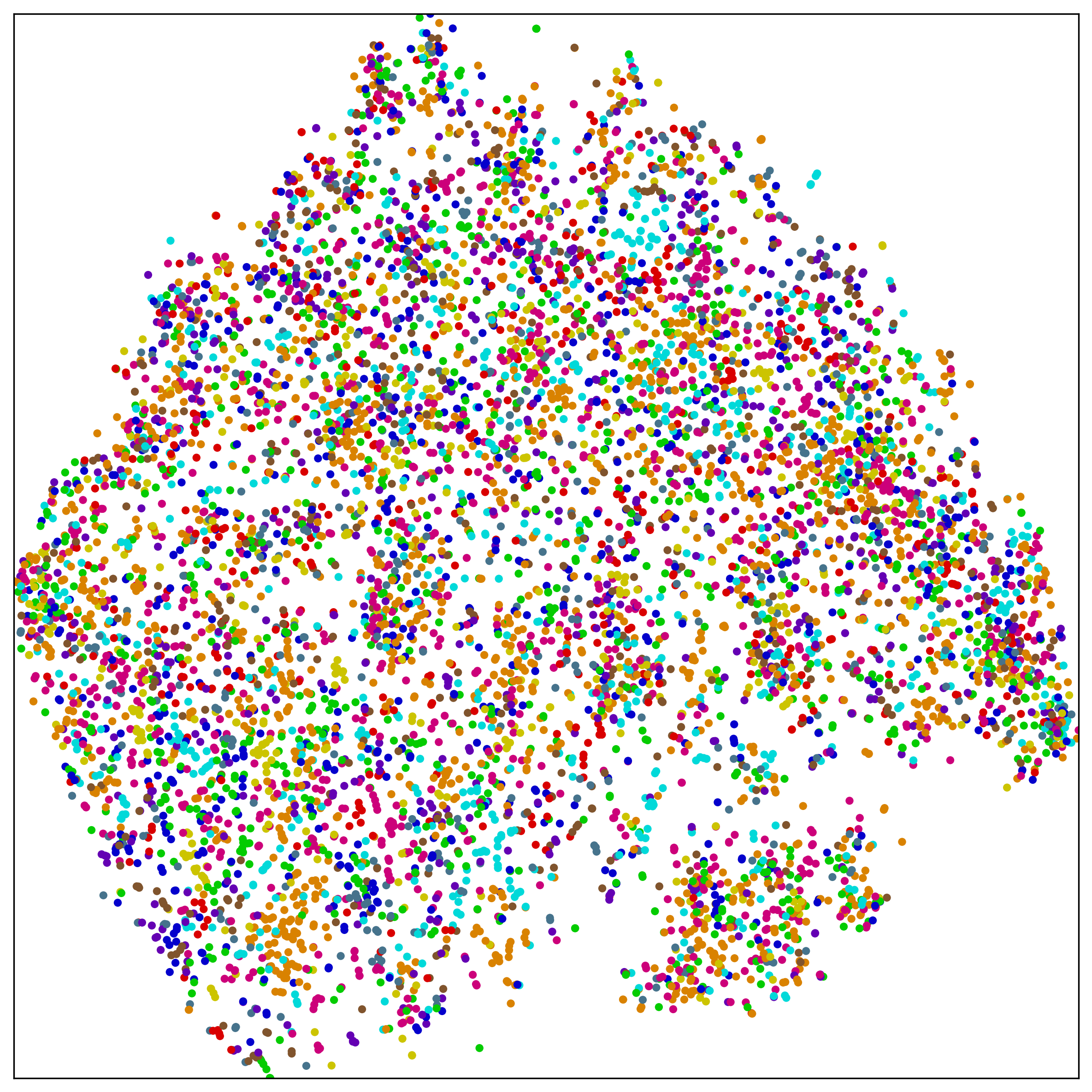}
            \includegraphics[width=0.45\linewidth]{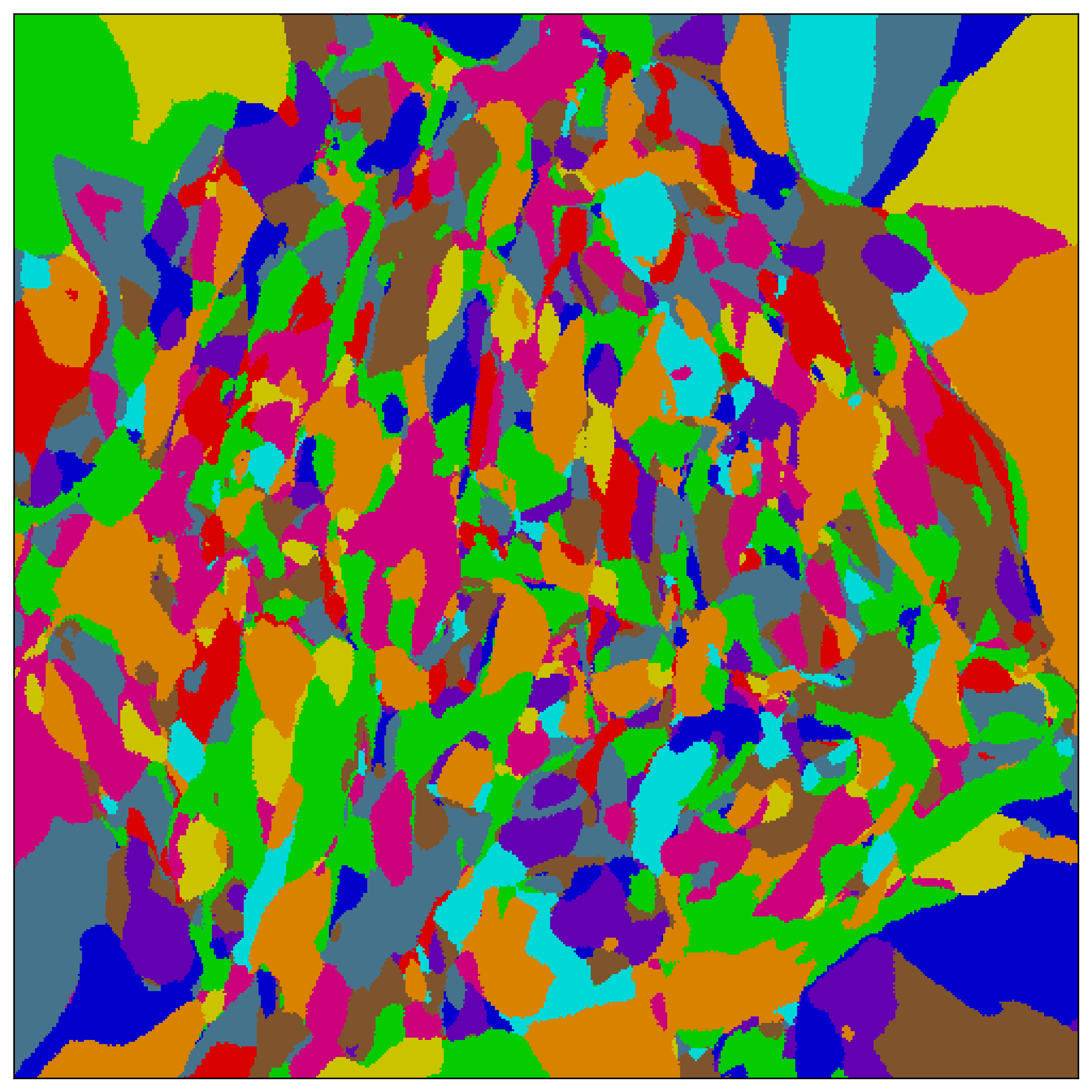}
            \includegraphics[width=0.04\linewidth]{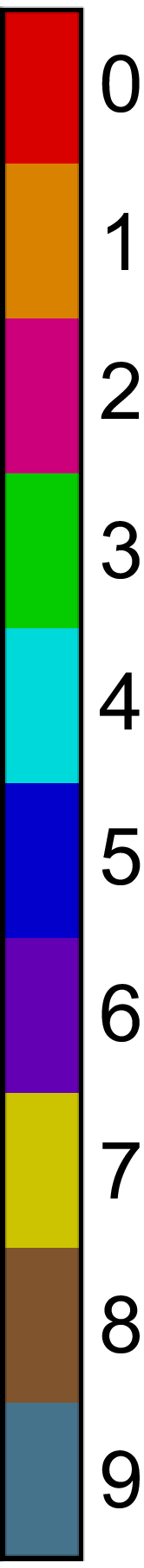}
            \caption{SVHN boundary map constructed using data projections $P(D)$.} 
            \label{subfig:svhn_subset_data}
        \end{subfigure}
        \begin{subfigure}[t]{\linewidth}
            \centering
            \includegraphics[width=0.45\linewidth]{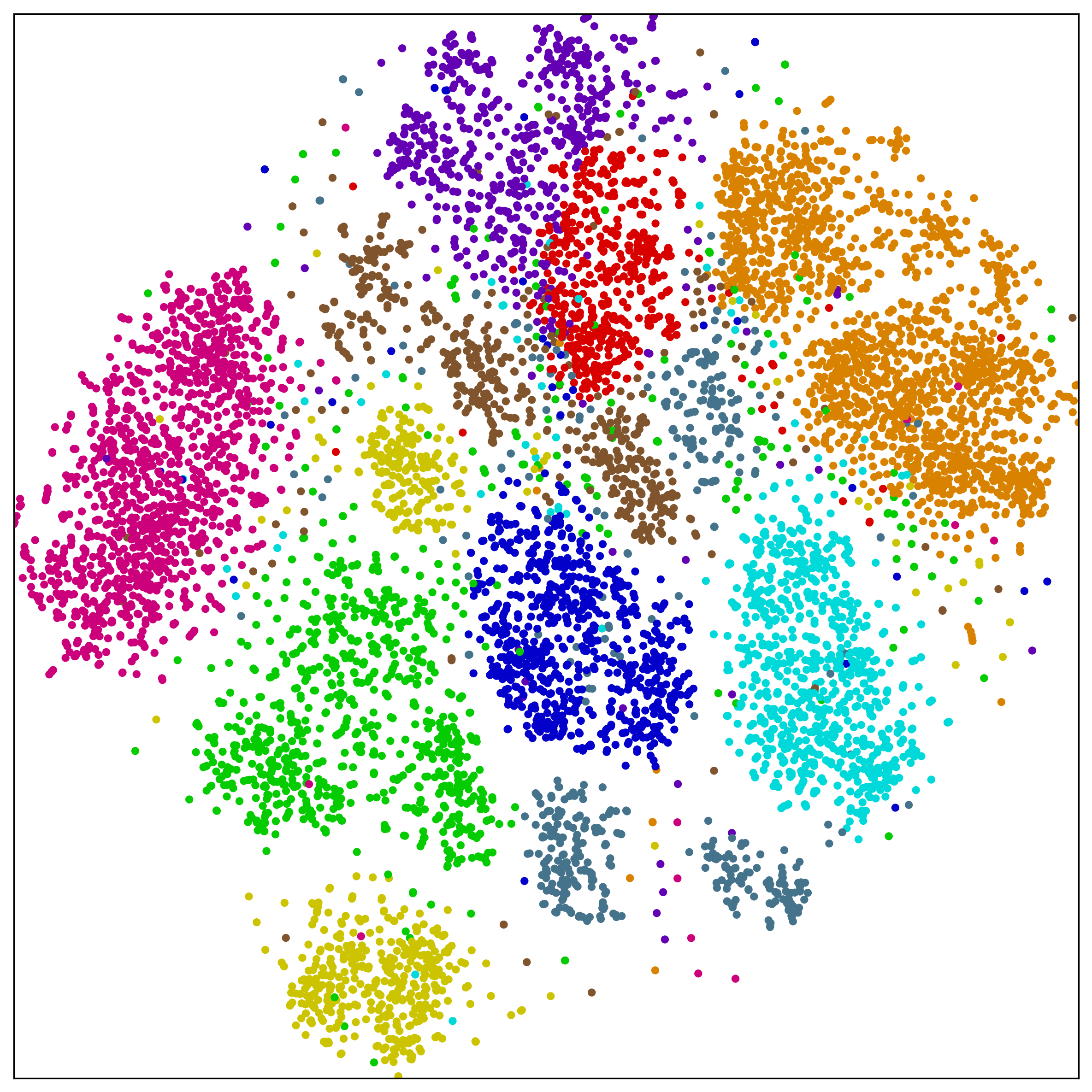}
            \includegraphics[width=0.45\linewidth]{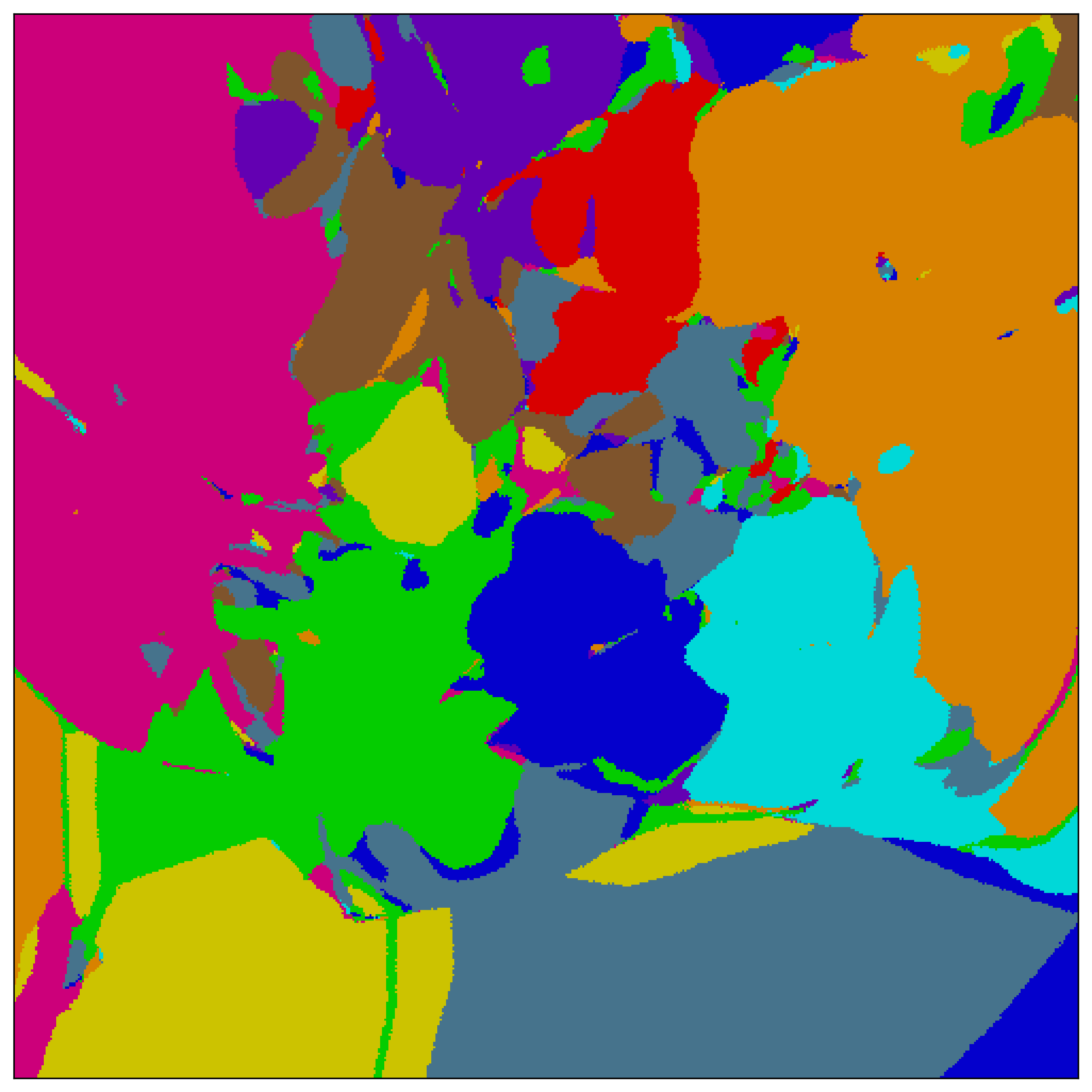}
            \includegraphics[width=0.04\linewidth]{cmap.png}
            \caption{SVHN boundary map constructed using Shapley value projections $P(S)$.} 
            \label{subfig:svhn_subset_shap}
        \end{subfigure}
        \caption{Scatterplots and DBMs for a 10k subset of SVHN using (a) Data space projections $P(D)$. (b) Shapley space projections $P(S)$.}
        \label{fig:svhn_subset}
    \end{figure}

    \textbf{Context:} As discussed in Section 6, Shapley value estimation for an entire dataset is computationally expensive. To prove that a subset is sufficient to produce a high-quality DBM, we use a subset of SVHN with 10k samples. Figure \ref{fig:svhn_subset} shows the resulting maps derived from data space projections (Figure \ref{subfig:svhn_subset_data}, $ACC_M = 28.7\%$) and Shapley value projections (Figure \ref{subfig:svhn_subset_shap}, $ACC_M = 86.2\%$).
    \newpage

\section{ShapDBMs using UMAP}
\begin{figure}[H]
        \begin{subfigure}[t]{\linewidth}
            \centering
                \includegraphics[width=0.45\linewidth]{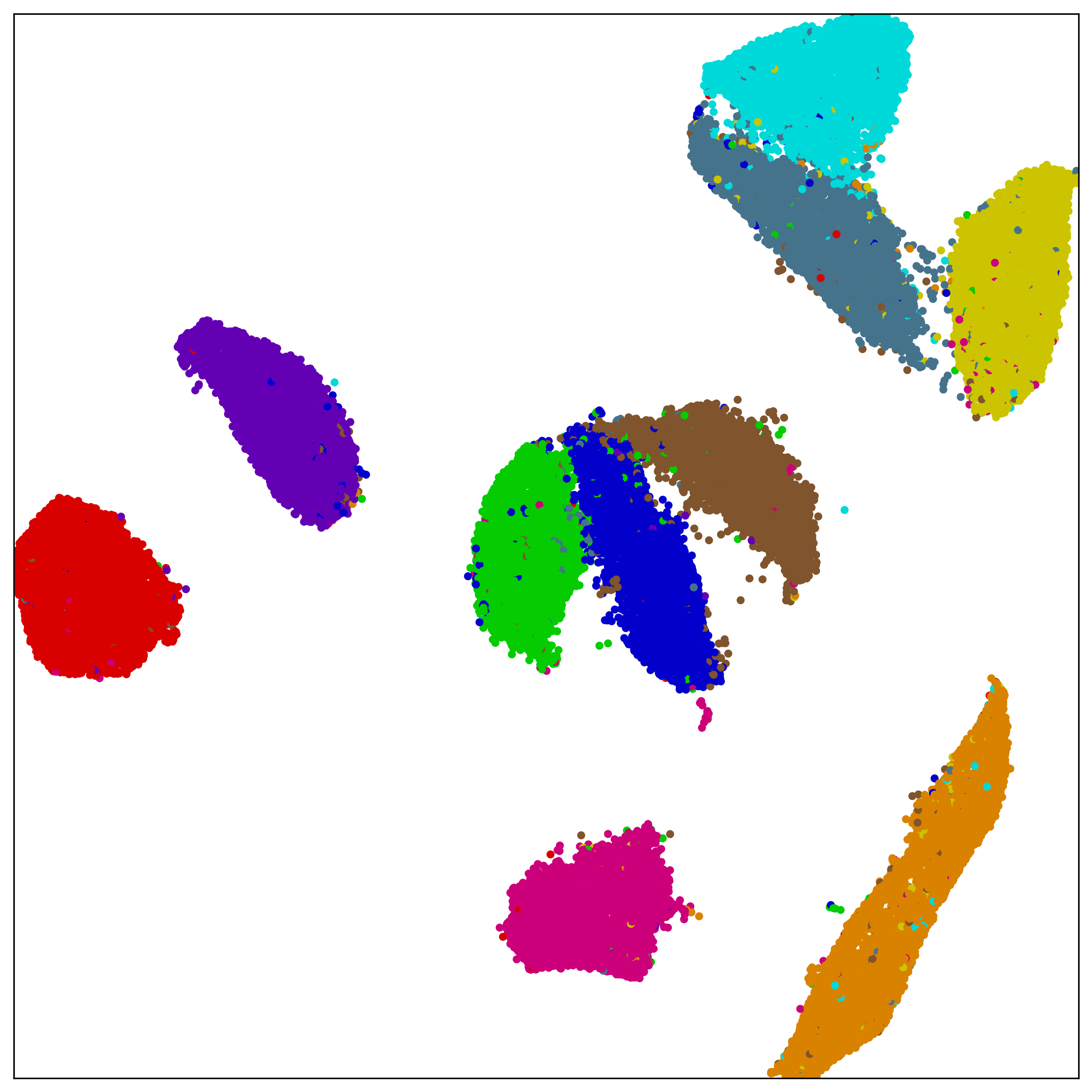}
                \includegraphics[width=0.45\linewidth]{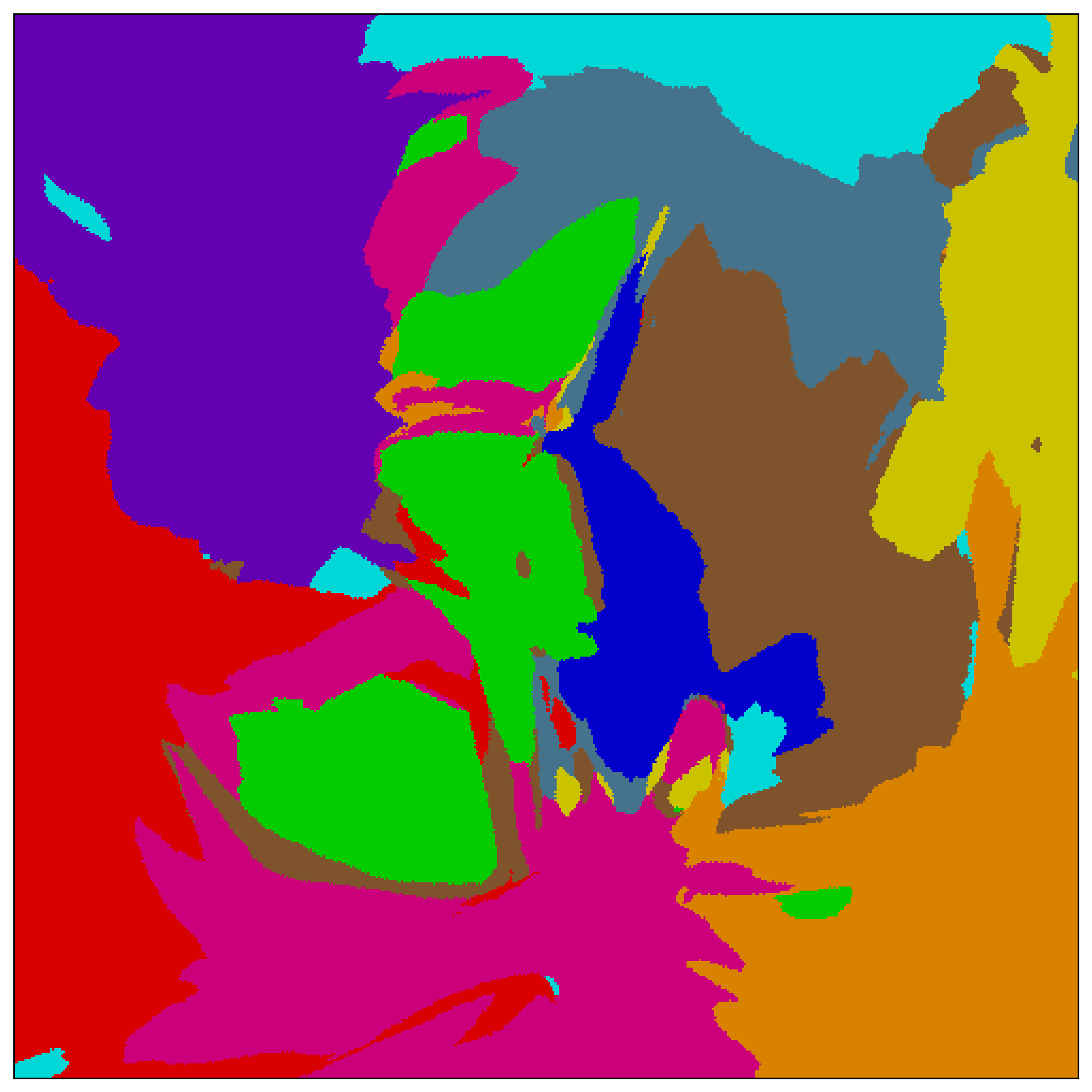}
                \includegraphics[width=0.04\linewidth]{cmap.png}
            \caption{MNIST boundary map using data projections $P(D)$.} 
            \label{subfig:mnist_data_umap}
        \end{subfigure}
        \begin{subfigure}[t]{\linewidth}
            \centering
            \includegraphics[width=0.45\linewidth]{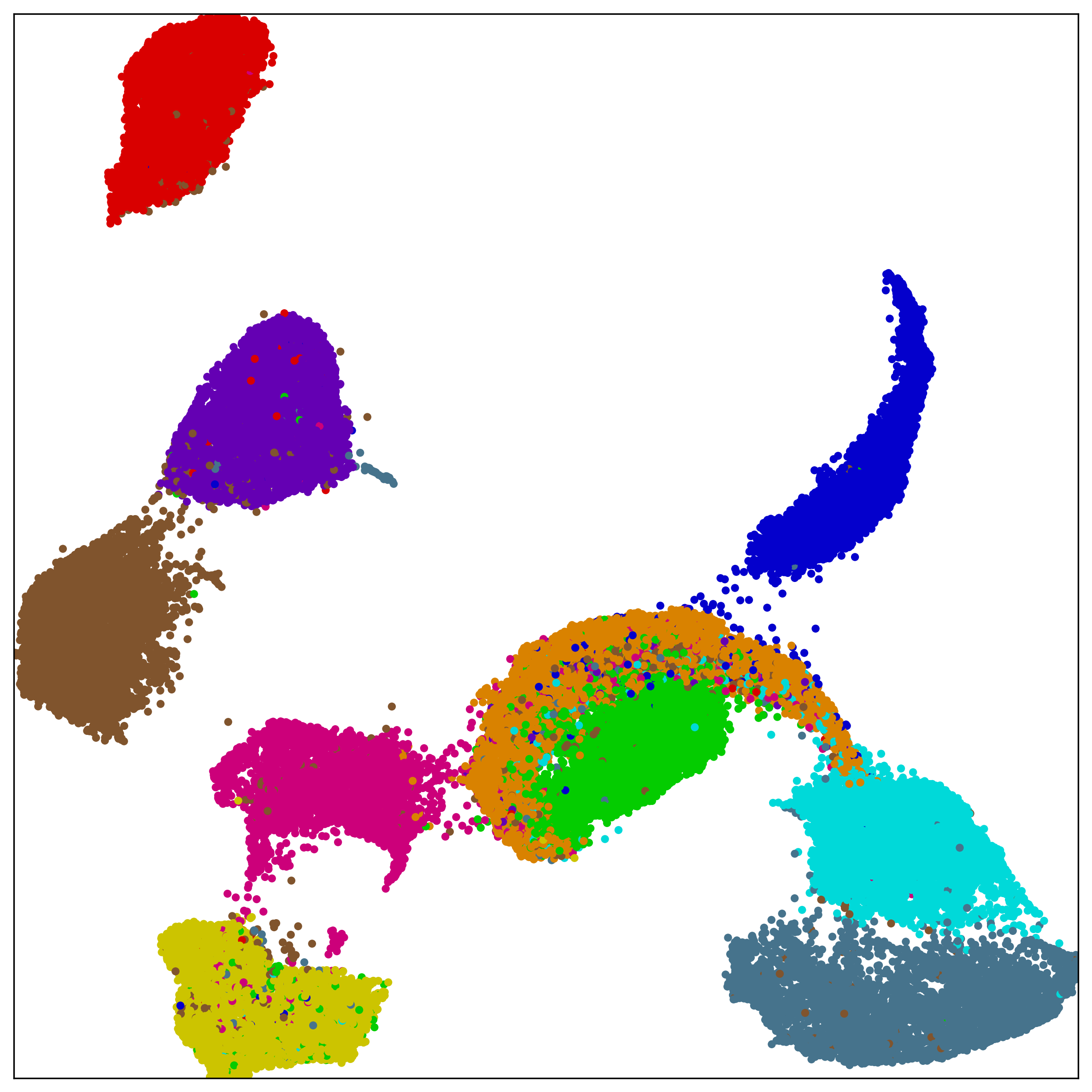}
            \includegraphics[width=0.45\linewidth]{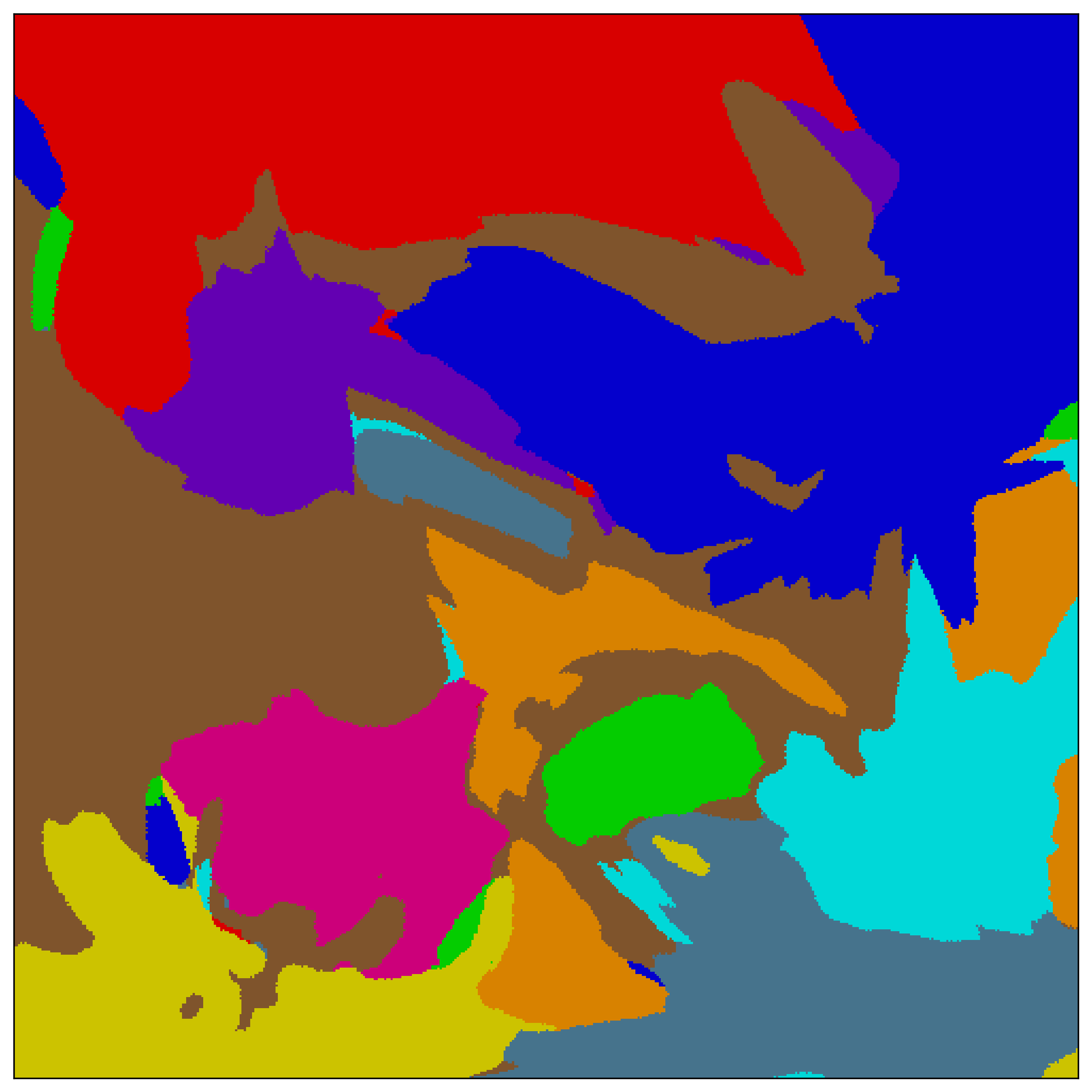}
            \includegraphics[width=0.04\linewidth]{cmap.png}
            \caption{MNIST boundary map using Shapley value projections $P(S)$.} 
            \label{subfig:mnist_shap_umap}
        \end{subfigure}
        \caption{Scatterplots and DBMs for MNIST using UMAP.  (a) Data space projections $P(D)$. (b) Shapley space projections $P(S)$.}
        \label{fig:mnist_umap}
        \vspace{-0.5cm}
    \end{figure}

\begin{figure}[H]
        \begin{subfigure}[t]{\linewidth}
            \centering
            \includegraphics[width=0.45\linewidth]{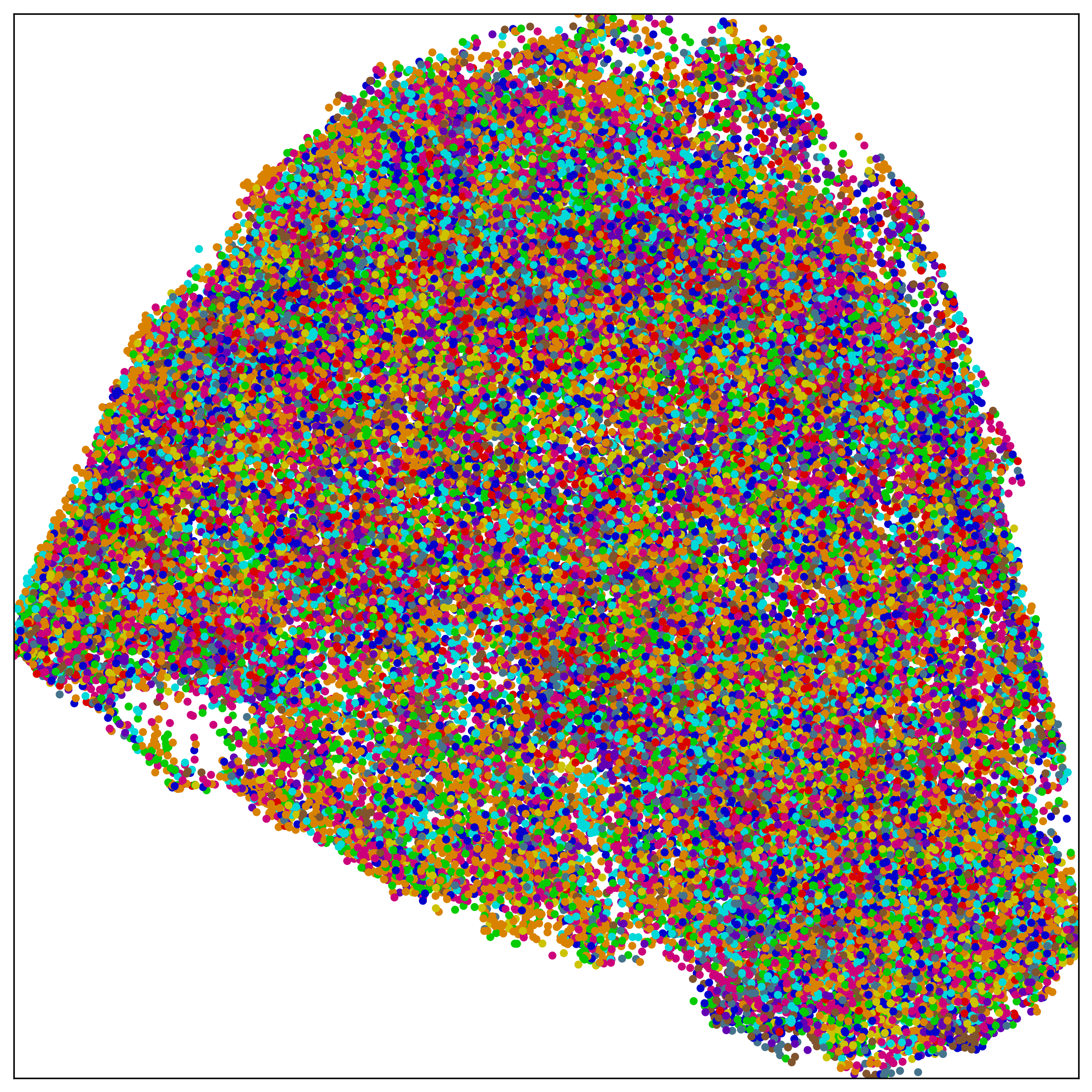}
            \includegraphics[width=0.45\linewidth]{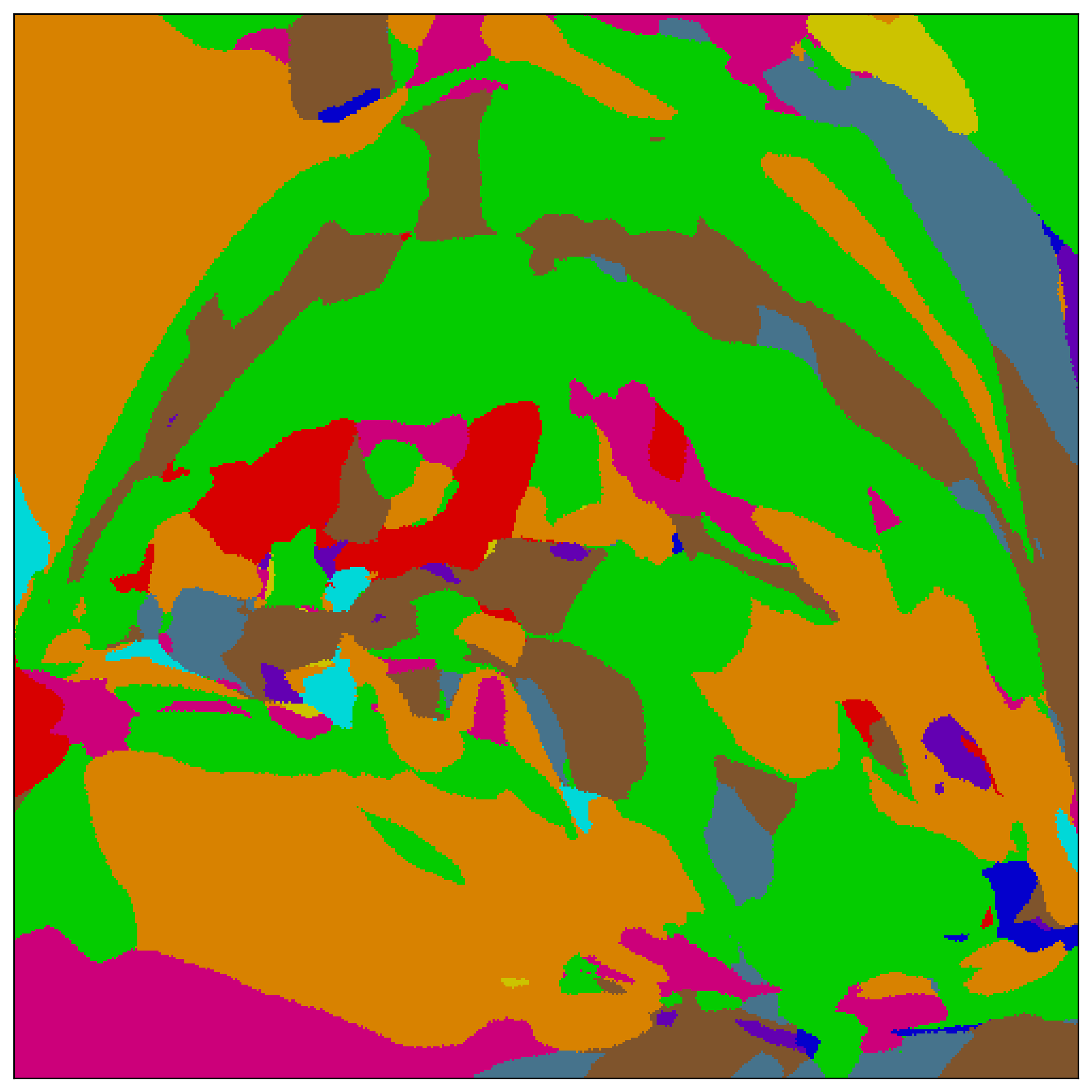}
            \includegraphics[width=0.04\linewidth]{cmap.png}
            \caption{SVHN boundary map using data projections $P(D)$.} 
            \label{subfig:mnist_data_umap}
        \end{subfigure}
        \begin{subfigure}[t]{\linewidth}
            \centering
            \includegraphics[width=0.45\linewidth]{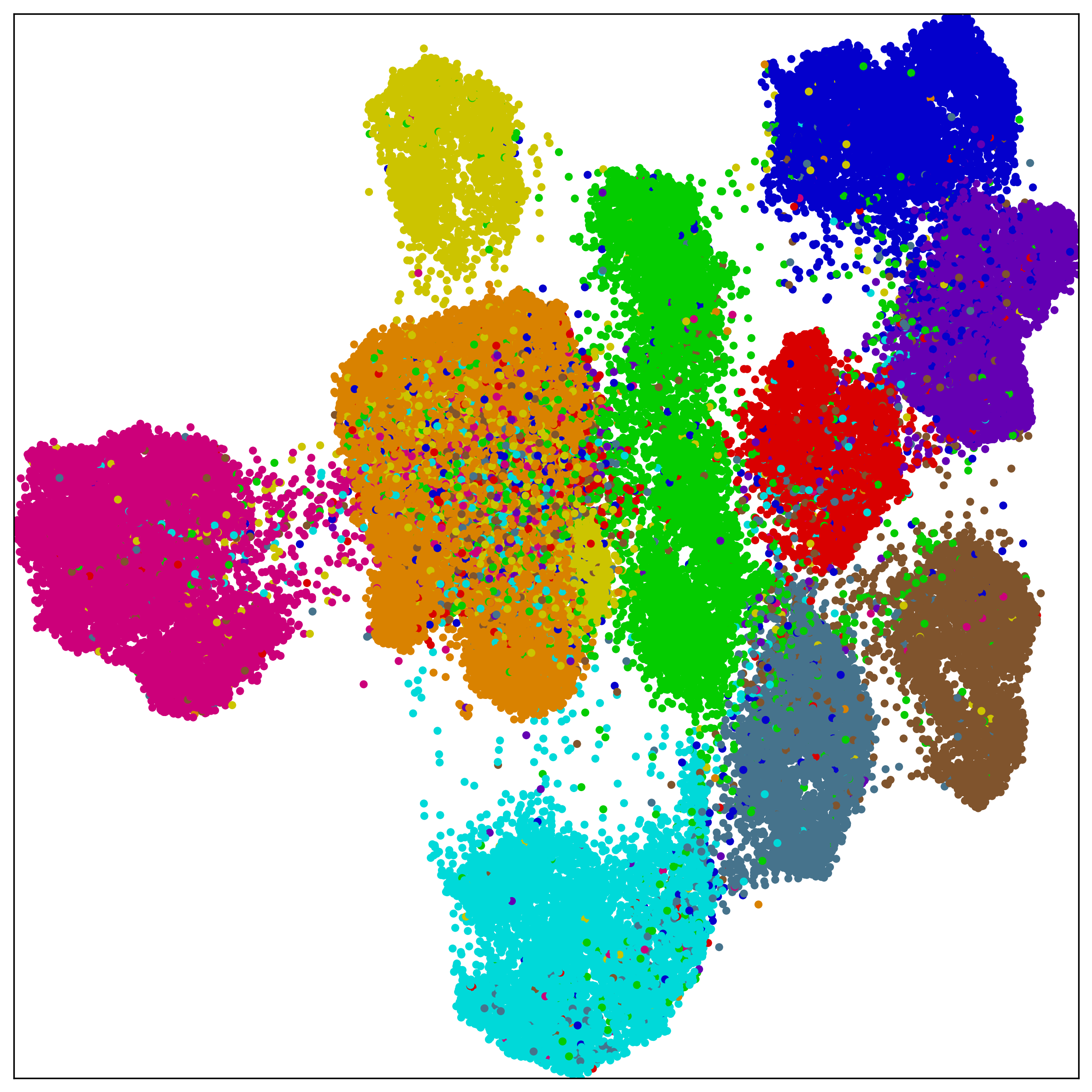}
            \includegraphics[width=0.45\linewidth]{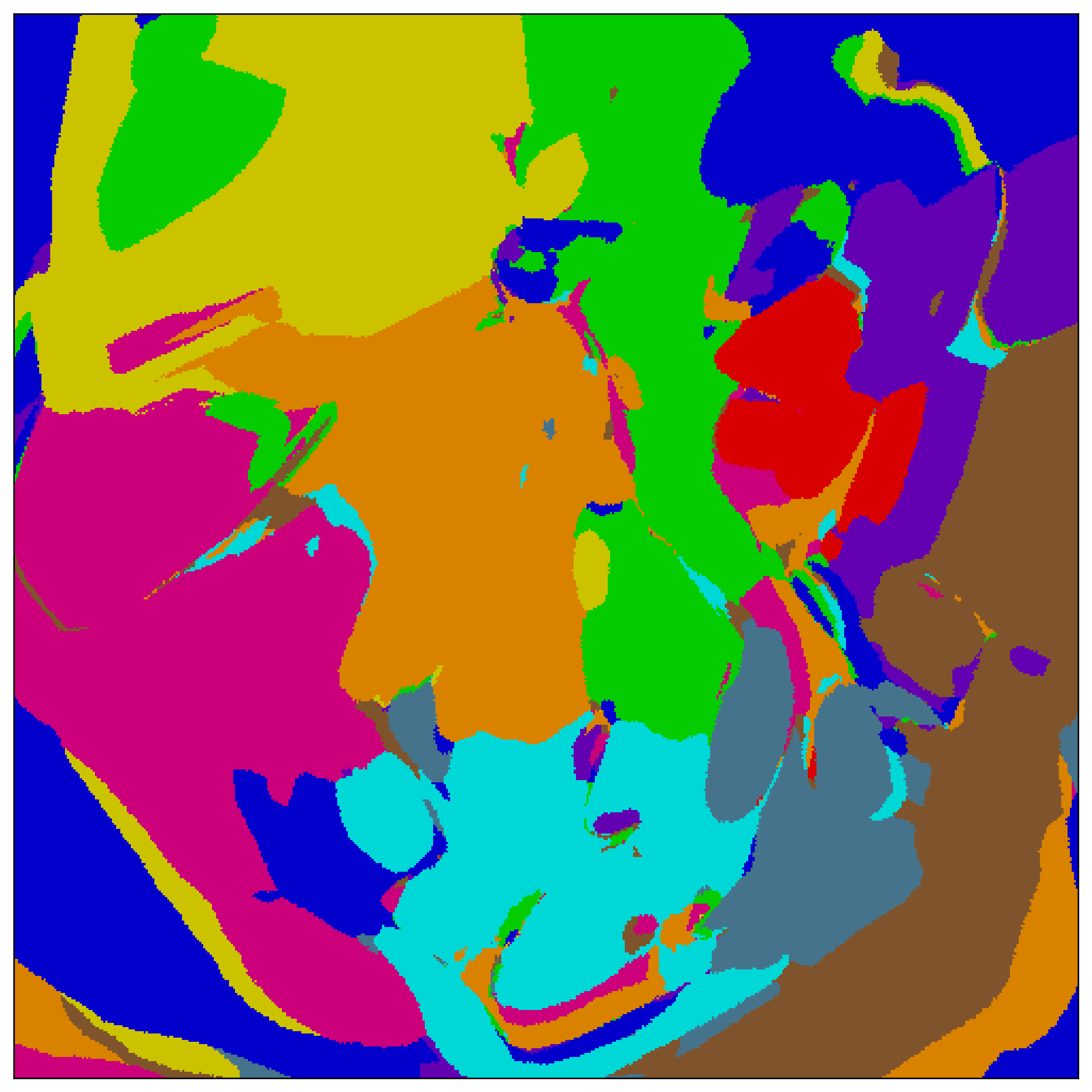}
            \includegraphics[width=0.04\linewidth]{cmap.png}
            \caption{SVHN boundary map using Shapley value projections $P(S)$.} 
            \label{subfig:mnist_shap_umap}
        \end{subfigure}
        \caption{Scatterplots and DBMs for SVHN using UMAP.  (a) Data space projections $P(D)$. (b) Shapley space projections $P(S)$.}
        \label{fig:svhn_umap}
        \vspace{-0.5cm}
    \end{figure}

\begin{figure}[H]
        \begin{subfigure}[t]{\linewidth}
            \centering
            \includegraphics[width=0.42\linewidth]{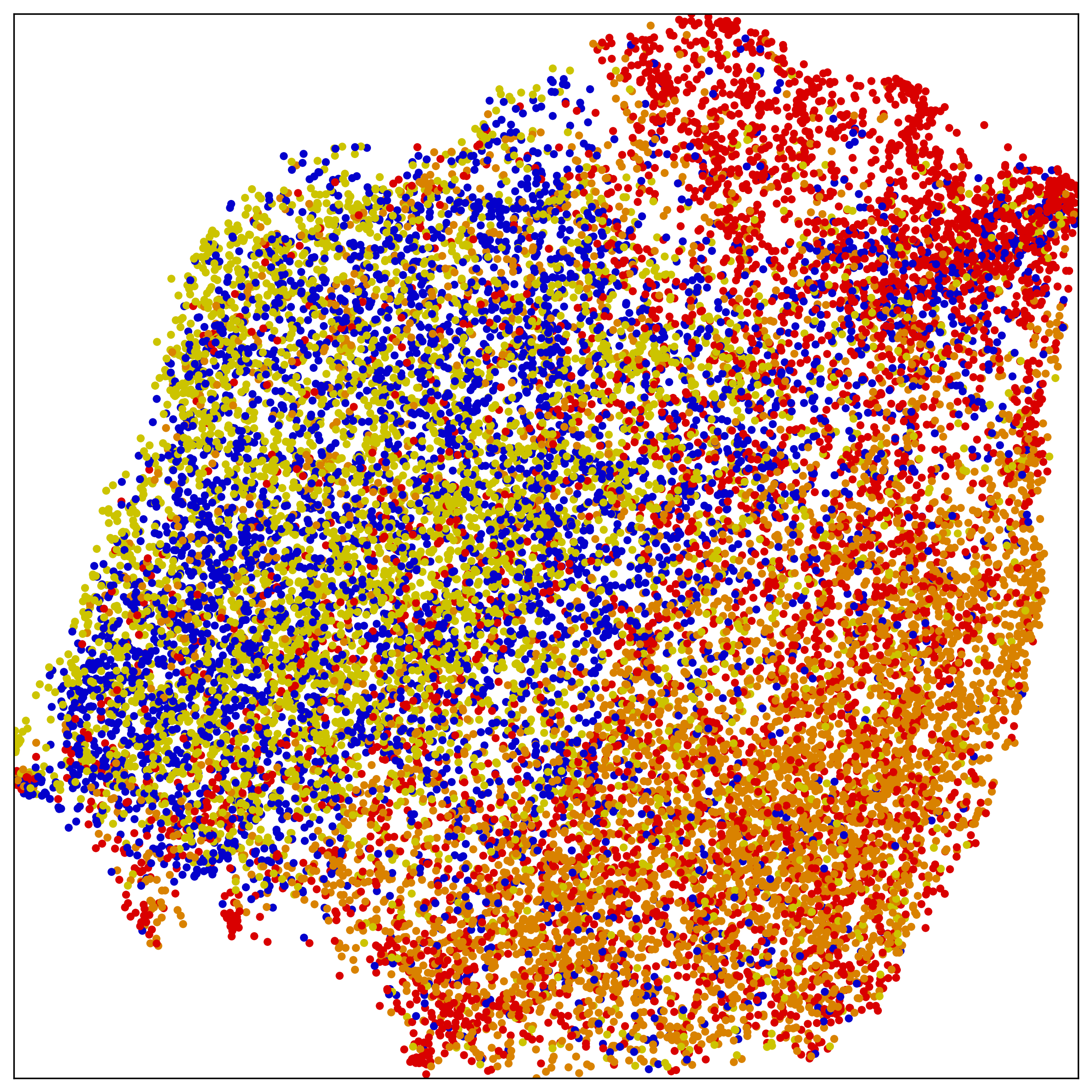}
            \includegraphics[width=0.42\linewidth]{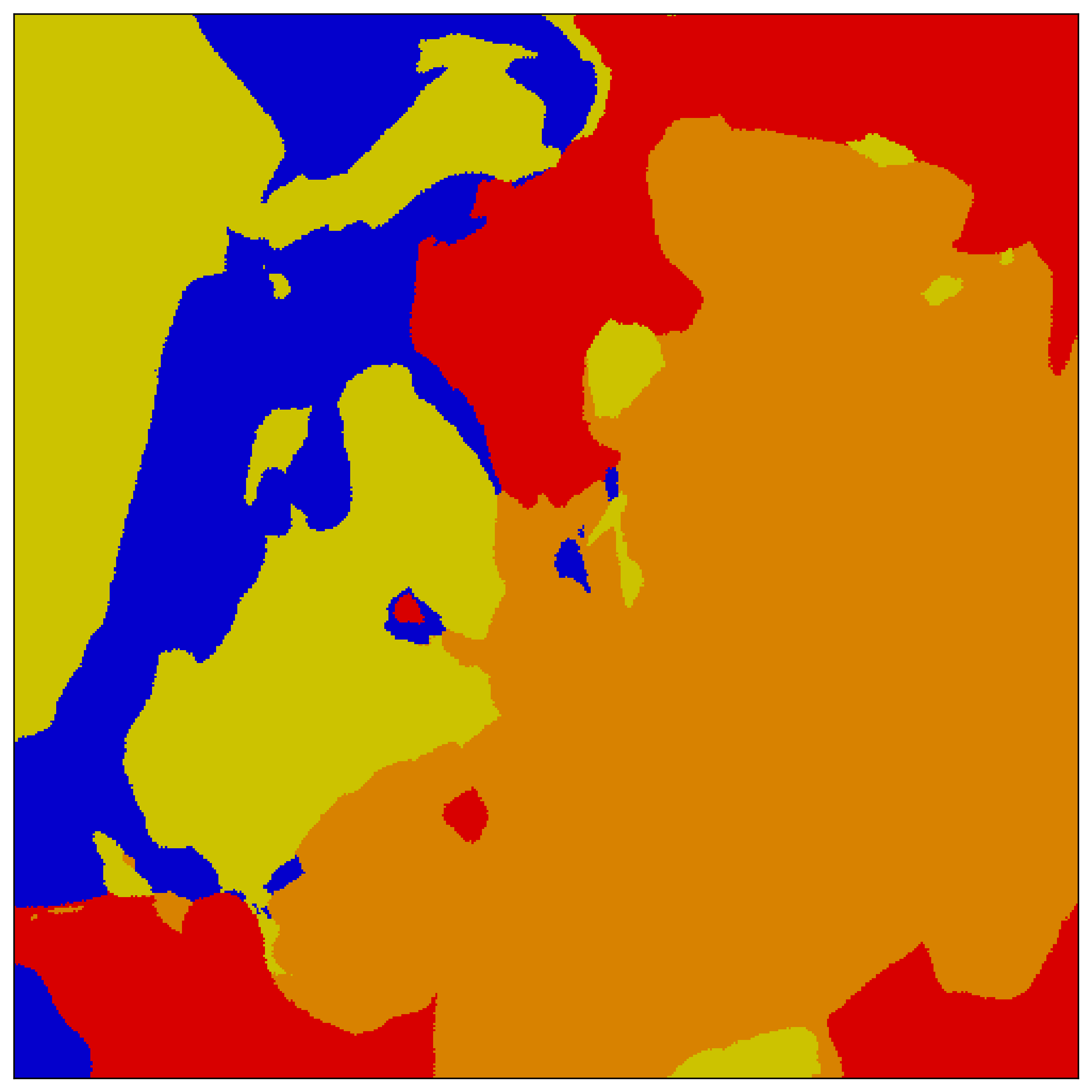}
            \includegraphics[width=0.09\linewidth]{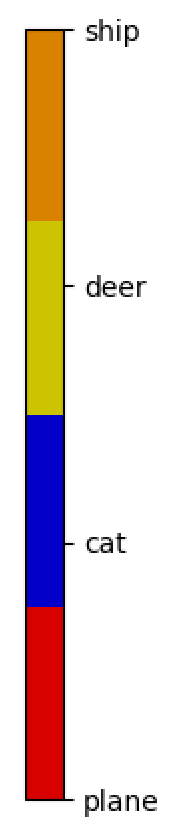}
            \caption{CIFAR-4 boundary map using data projections $P(D)$.} 
            \label{subfig:mnist_data_umap}
        \end{subfigure}
        \begin{subfigure}[t]{\linewidth}
            \centering
            \includegraphics[width=0.42\linewidth]{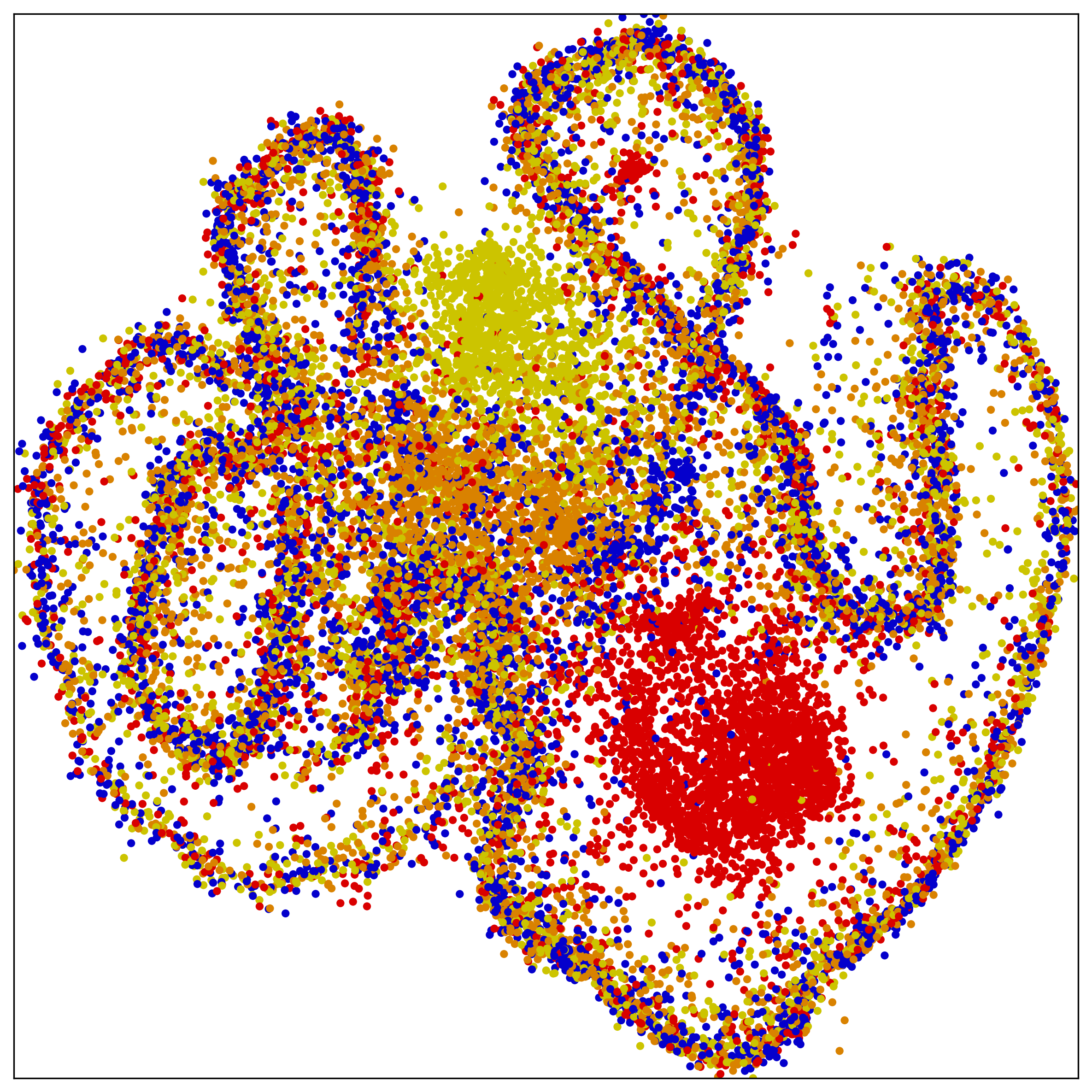}
            \includegraphics[width=0.42\linewidth]{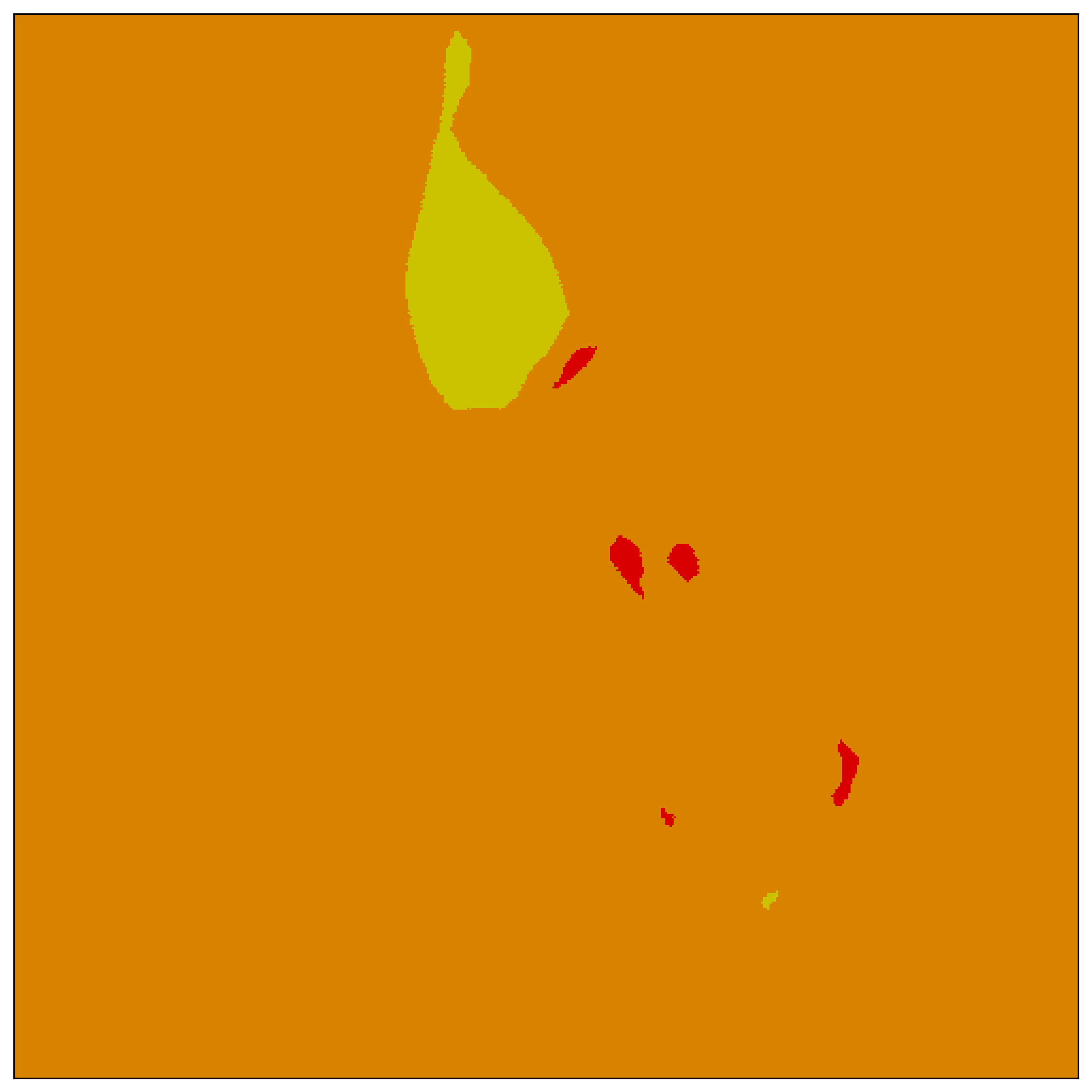}
            \includegraphics[width=0.09\linewidth]{cmap_4_vert.png}
            \caption{CIFAR-4 boundary map using Shapley value projections $P(S)$.} 
            \label{subfig:mnist_shap_umap}
        \end{subfigure}
        \caption{Scatterplots and DBMs for CIFAR-4 using UMAP.  (a) Data space projections $P(D)$. (b) Shapley space projections $P(S)$.}
        \label{fig:cifar_umap}
        \vspace{-0.5cm}
    \end{figure}

    \textbf{Context:} As discussed in section 4, all DBMs are calculated using t-SNE projections. Figures \ref{fig:mnist_umap}, \ref{fig:svhn_umap} and \ref{fig:cifar_umap} show the DBMs based on data space and Shapley space UMAP projections for MNIST, SVHN and CIFAR-4 respectively.

\end{document}